\documentclass[amsmath,amssymb,10pt,aps,notitlepage,prl,twocolumn,nofootinbib,longbibliography]{revtex4-1} 
 \usepackage[colorlinks=true]{hyperref}
 \usepackage{lipsum}
 \usepackage{graphicx}
\usepackage{latexsym}
\usepackage{amsmath}
\usepackage{amsthm}
\usepackage{amssymb}
\usepackage{epstopdf} 
\usepackage{enumitem}
\usepackage{setspace}
\usepackage{dcolumn}
\usepackage{bm}
\usepackage{setspace} 
\usepackage{slashed}
\usepackage{color}
\usepackage{youngtab}
\usepackage{tikz}
\usepackage{tikz-3dplot}
\usepackage{braket}
\usepackage{array}
\usepackage{tabularx}
\usepackage{multirow}  
\usepackage{makeidx}
\usepackage{cancel}
\usepackage{subfigure}
\usepackage{pbox}
\usepackage{pifont}
\usepackage{newunicodechar}
\usepackage{mathrsfs}  
\usepackage{siunitx}
\usepackage[utf8]{inputenc}
\usepackage{romanbar}

\usepackage{lipsum}
\graphicspath{{Figures/}}

\usepackage{lipsum}
\graphicspath{{Figures/}}

\usetikzlibrary{shapes,snakes,arrows,chains,matrix,positioning,scopes,calc}
\usetikzlibrary{decorations.markings}

 \usepackage{moreverb}
\allowdisplaybreaks 
\begin{document}

\title{Fully Generalized Spin Models with Strain Effects \\ of Kitaev Spin Liquid Candidate Materials}

\author{Pureum Noh}
\thanks{These authors contributed equally: Pureum Noh, Hyunggeun Lee.}
 \affiliation{Department of Physics, Korea Advanced Institute of Science and Technology (KAIST), Daejeon 34141, Korea}

\author{Hyunggeun Lee}
\thanks{These authors contributed equally: Pureum Noh, Hyunggeun Lee.} 
\affiliation{Department of Physics, Korea Advanced Institute of Science and Technology (KAIST), Daejeon 34141, Korea}

 \author{Myung Joon Han}
\thanks{mj.han@kaist.ac.kr}
\affiliation{Department of Physics, Korea Advanced Institute of Science and Technology (KAIST), Daejeon 34141, Korea}

 \author{Eun-Gook Moon}
\thanks{egmoon@kaist.ac.kr}
\affiliation{Department of Physics, Korea Advanced Institute of Science and Technology (KAIST), Daejeon 34141, Korea}

{
\begin{abstract}
The $KJ\Gamma\Gamma'$ spin model—originally derived for an ideal $P\bar{3}1m$ symmetric geometry—has long served as a central framework for understanding candidate Kitaev materials. In realistic crystals, however, this ideal geometry is seldom realized, either at low temperatures or under external perturbations, limiting the model’s quantitative applicability. Here we introduce a fully generalized spin model, denoted $\epsilon$-$KJ\Gamma\Gamma'$, that explicitly incorporates arbitrary lattice deformations $\epsilon$. 
All spin-exchange interactions and their strain-dependent coefficients are obtained from density-functional theory (DFT) calculations and a microscopic derivation of coupling constants for materials based on $d^5$ transition-metal ions. For $\alpha$-RuCl$_3$ under a strain of $3\%$, new emergent exchange channels acquire magnitudes comparable to their unstrained counterparts. Building on these parameters, we investigate strain-driven quantum phase transitions between competing magnetic states—including the zigzag order and the Kitaev quantum spin liquid (KQSL)—and identify a strain-induced topological transition within the KQSL states that offers a practical diagnostic of Kitaev physics. 
Furthermore, our symmetry analysis of the $\epsilon$-$KJ\Gamma\Gamma'$ model is applicable to both $d^{5}$ ions, such as $\alpha$-RuCl$_3$, and $d^{7}$ systems, including cobalt-based compounds.

 
\end{abstract}

\maketitle

Quantum spin liquid (QSL) states have gained significant attention in future science and technology research due to their inherent abundance of quantum entanglement, which inhibits the formation of conventional magnetic ordering and gives rise to the emergence of exotic excitations
\cite{Zhou2017.4,Savary2016.11,Balents2010.3,Anderson1973.2,Knolle2019.3}.
Among various QSL states, the Kitaev quantum spin liquid (KQSL) state 
under a magnetic field exhibits a non-trivial topological invariant and hosts non-Abelian anyon excitations, making it
a promising topological computation platform \cite{Nayak2008.9, Kitaev2002.5,Kitaev2005.10}.
A realization of the KQSL has been proposed in materials based on $d^{5}$ ions such as honeycomb iridates and $\alpha$-RuCl$_3$, where microscopic calculations suggest that their interactions can be well described by a pure Kitaev model for $j_{\text{eff}}=1/2$ spins mediated through edge-shared octahedra \cite{Jackeli2009.1}. 
%
In the $P\bar{3}1m$ symmetric structure, microscopic calculations have successfully explained the generic nearest-neighbor spin model, known as the $KJ\Gamma \Gamma'$ model, which extends the pure Kitaev model by incorporating additional exchange interactions \cite{Chaloupka2010.7,Rau2014.2,rau_trigonal_2014}.
Extensive studies of the $KJ\Gamma\Gamma'$ model have revealed a variety of fascinating phenomena, including an extended spin liquid phase and quantum phase transitions into several well-understood magnetic ground states \cite{Plumb2014.7,Koitzsch2016.9,Sandilands2015.4,Kim2015.6,Kim2016.4,Winter2017.11,Winter2016.6,Winter2018.2,Yadav2016.11,Takeda2022.11,Viciu2007.1,Songvilay2020.12,Lin2021.9,Wulferding2020.3,Tanaka2022.1,Xing2025.3,Kasahara2022.8,Kasahara2018.7,Yokoi2021.7,Czajka2021.5,Takagi2019.3,Lefrancois2022.4,Rousochatzakis2015.12,Halasz2016.9,Ronquillo2019.4,Patel2019.5,Vinkler-Aviv2018.8,Knolle2014.5,Nasu2016.7,Nasu2015.9,Nasu2017.9,Motome2019.12,Kasahara2018.5,Yamada2017.8,Gohlke2017.10,Cookmeyer2023.6,Go2019.4,Yang2022.8, Halasz.G.B.2025.05,Zhang.H.2025.05,Zhang.E.Z.2023.5,Halloran.T.2022.11}, most notably providing a compelling explanation for the zigzag ground state observed in $\alpha$-RuCl$_3$ \cite{Rau2014.2}.

However, the symmetry in real materials is lower than that of the $P\bar{3}1m$ symmetric structure, indicating that the $KJ\Gamma\Gamma'$ microscopic theory is not sufficient for a complete description.
Experimentally, not only the $P\bar{3}1m$ symmetric structure but also some lower-symmetry structures
have been reported for $\alpha$-RuCl$_3$ \cite{Fletcher1967.1,Kubota2015.3}.
To incorporate the effect of symmetry lowering, bond-dependent $K, J,\Gamma,\Gamma'$ have been introduced within the framework of $KJ\Gamma\Gamma'$ model \cite{Kim2016.4,Kaib2021.4}.
%
This correction is limited both quantitatively and qualitatively, as it cannot account for all possible additional spin interactions due to insufficient degrees of freedom.
%
Therefore, a more general microscopic approach is strongly required to fully describe real materials and to investigate the effect of generic lattice deformations on the materials.

\begin{figure*}[tb]
\includegraphics[width=0.7\textwidth]{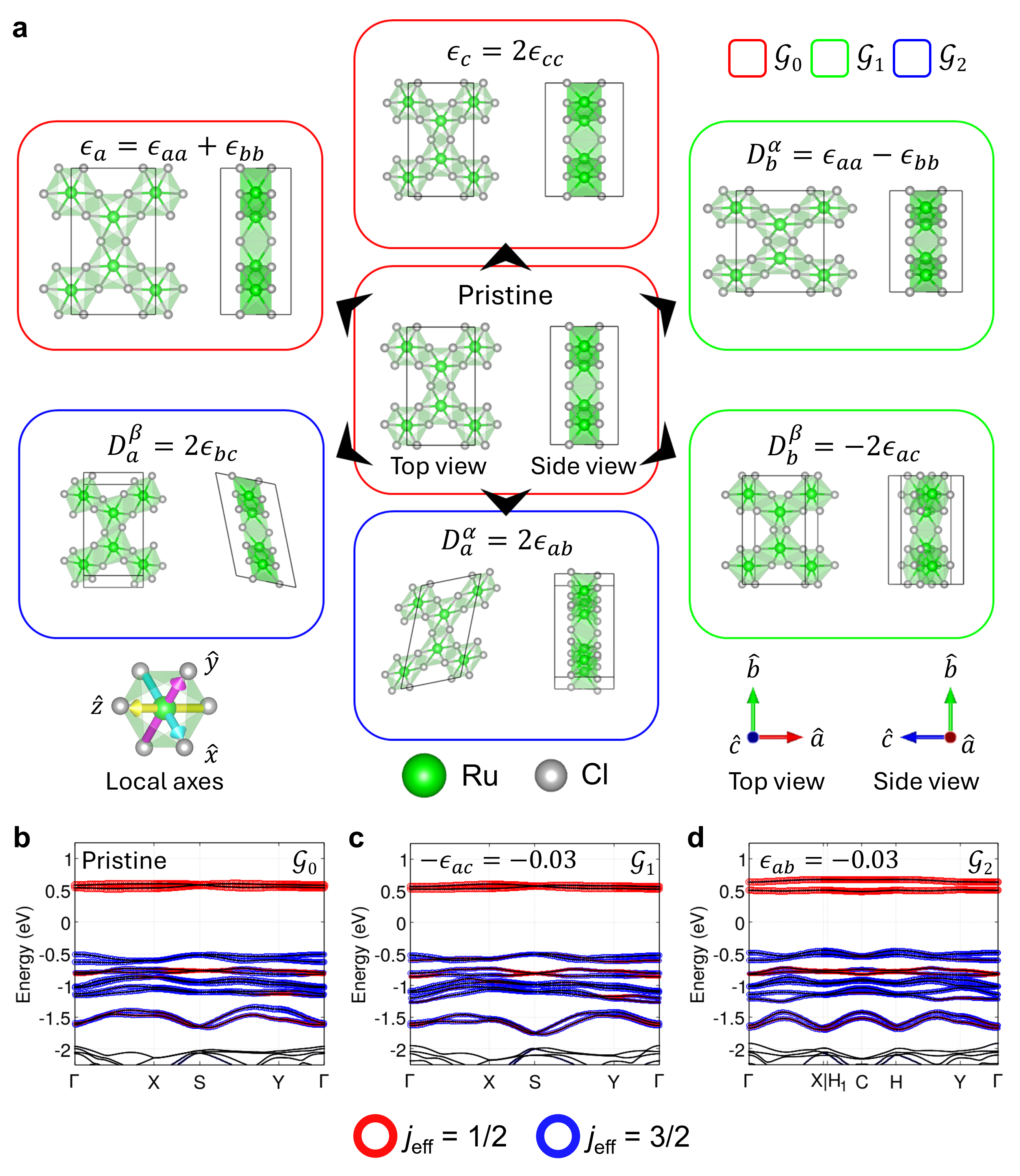}
\caption{The crystal structures and the $j_\text{eff}$-projected band dispersions of strained $\alpha$-RuCl$_3$ monolayer. \textbf{a}, Crystal structures of the pristine (middle) and the deformed RuCl$_3$ under six independent strains. Cyan, magenta, and yellow arrows in the left bottom panel represent the local $x$-, $y$-, and $z$-axes, respectively. Red, green, and blue boxes indicate the categorized group of $\mathcal{G}_0$, $\mathcal{G}_1$, and $\mathcal{G}_2$, sequentially. \textbf{b}-\textbf{d}, The calculated electronic structures of \textbf{b} pristine, \textbf{c} $D^{\beta}_b$-, and \textbf{d} $D^{\alpha}_a$-strained RuCl$_3$ monolayer by $-3$\%. The size of red and blue circles indicate the weight of $j_\text{eff}=1/2$ and $3/2$ character, respectively.}
\label{fig1}
\end{figure*}

Here, we provide a generalized spin model with explicit parameter values for homogeneous strains through microscopic analysis and density functional theory (DFT) calculations.
%
%
%
%
After classifying the symmetry of strains, we identify strains which preserve the forms of the conventional $KJ\Gamma\Gamma'$ model while the relevant coupling constants are a function of such strains. 
Such strains may be utilized to stabilize KQSLs.
The other types of strains induce additional interaction terms beyond the $KJ\Gamma\Gamma'$ model since the $P\bar{3}1m$ symmetry of the $KJ\Gamma\Gamma'$ model is broken. We refer to the resulting extended model as the $\epsilon$-$KJ\Gamma\Gamma'$ model.
Our DFT results indicate that the strengths of the newly introduced interactions are comparable to those in the original $KJ\Gamma\Gamma'$ model. This highlights the potential of the $\epsilon$-$KJ\Gamma\Gamma'$ model as a valuable framework for investigating diverse quantum phase transitions in a manner analogous to the role played by $KJ\Gamma\Gamma'$ model in previous studies.
As one of the quantum phase transitions, we have identified a topological quantum phase transition associated with the mass of Majorana fermions in the KQSL, which can be controlled through strain even in the presence of a bond-directional magnetic field, where massless Majorana fermions are known to emerge.
While our DFT calculations focused on $\alpha$-RuCl$_3$, our microscopic analysis is directly applicable to other materials based on $d^5$ ions. Moreover, our symmetry analysis of the $\epsilon$-$KJ\Gamma\Gamma'$ model, which clarifies the coupling constants' dependence on the strain directions, is also applicable to $d^7$ systems, including cobalt-based compounds.

\begin{table*}[tb]
\centering 
\begin{tabular}{>{\centering}p{3cm}|>{\centering\arraybackslash} p{13cm}}
\hline 
\hline 
Our notation& Strain  \\ 
\hline \hline 
$\epsilon_a$ & $\epsilon_{aa}+\epsilon_{bb}$ \\
$\epsilon_c$  & $2\epsilon_{cc}$  \\
\hline
\normalsize{$\mathbf{D}^{\alpha}=(D^{\alpha}_a,D^{\alpha}_b)$} & $(2\epsilon_{ab},\epsilon_{aa}-\epsilon_{bb})$ \\
\normalsize{$\mathbf{D}^{\beta}=(D^{\beta}_a,D^{\beta}_b)$} & $(2\epsilon_{bc},-2\epsilon_{ac})$  \\
\hline 
\hline 
Our notation& Nearest neighbor's two spins \\ 
\hline \hline 
$\overrightarrow{\mathcal{S}}^{K_{(\mathcal{E})}}$ & \normalsize{$\frac{1}{2\sqrt{6}}\left(\sqrt{3}\left[J^{xx}_x-J^{yy}_y\right],\;J^{xx}_x+J^{yy}_y-2J^{zz}_z\right)$}  \\
$\mathcal{S}^{J_{(\mathcal{A}_2)}}$ & $\frac{1}{2}(J^{xx}_{y}+J^{yy}_{z}+J^{zz}_{x}-J^{xx}_{z}-J^{yy}_{x}-J^{zz}_{y})$\\
$\overrightarrow{\mathcal{S}}^{J_{(\mathcal{E},1)}}$ & \normalsize{$\frac{1}{4\sqrt{3}}\left(J^{xx}_{y}+J^{yy}_{z}-2J^{zz}_{x}-J^{xx}_{z}+2J^{zz}_{y}-J^{yy}_{x},\;\sqrt{3}\left[-J^{xx}_{y}+J^{yy}_{z}+J^{xx}_{z}-J^{yy}_{x}\right]\right)$}\\
$\overrightarrow{\mathcal{S}}^{J_{(\mathcal{E},2)}}$ & \normalsize{$\frac{1}{4\sqrt{3}}\left( \sqrt{3}\left[J^{xx}_{y}-J^{yy}_{z}+J^{xx}_{z}-J^{yy}_{x}\right],\; J^{xx}_{y}+J^{yy}_{z}-2J^{zz}_{x}+J^{xx}_{z}-2J^{zz}_{y}+J^{yy}_{x} \right)$}  \\
$\overrightarrow{\mathcal{S}}^{\Gamma_{(\mathcal{E})}}$ & \normalsize{$\frac{1}{2\sqrt{3}}\left(\sqrt{3}\left[J^{yz}_{x}-J^{zx}_{y}\right],\; -2J^{xy}_{z}+J^{yz}_{x}+J^{zx}_{y} \right)$} \\
$\mathcal{S}^{\Gamma'_{(\mathcal{A}_2)}}$ & $J^{xy}_{x}+J^{yz}_{y}+J^{zx}_{z}-J^{xz}_{x}-J^{zy}_{z}-J^{yx}_{y}$\\
$\overrightarrow{\mathcal{S}}^{\Gamma'_{(\mathcal{E},1)}}$ & \normalsize{$\frac{1}{2\sqrt{6}}
\left(J^{xy}_{x}+J^{yz}_{y}-2J^{zx}_{z}-J^{xz}_{x}+2J^{zy}_{z}-J^{yx}_{y},\;\sqrt{3}\left[-J^{xy}_{x}+J^{yz}_{y}+J^{xz}_{x}-J^{yx}_{y}\right]\right)$}\\
$\overrightarrow{\mathcal{S}}^{\Gamma'_{(\mathcal{E},2)}}$ & \normalsize{$\frac{1}{2\sqrt{6}}
\left( \sqrt{3}\left[J^{xy}_{x}-J^{yz}_{y}+J^{xz}_{x}-J^{yx}_{y}\right],\;J^{xy}_{x}+J^{yz}_{y}-2J^{zx}_{z}+J^{xz}_{x}-2J^{zy}_{z}+J^{yx}_{y} \right)$}\\[0.1cm]
\hline
\hline
\end{tabular}
\caption{ All relevant irreducible representations for strains and nearest neighbor's two spins. $ J^{\alpha \beta}_{\gamma}=\sum_{\braket{j,k}_{\gamma}} (S^{\alpha}_{j}S^{\beta}_{k}+S^{\beta}_{j}S^{\alpha}_{k})$ where $\braket{j,k}_{\gamma}$ are for the nearest-neighbor bonds with a component $\gamma=x, y, z$.}
 \label{table1}
\end{table*}

\section*{Result}
\noindent\textbf{General Strategy.} In this section, we propose a general strategy for constructing a generalized spin model under external perturbations, providing an overview of the computational steps we performed.
We consider a system with a symmetry group $\mathcal{G}_0$ whose Hamiltonian is denoted as $H_0$. An external symmetry-breaking perturbation with $\epsilon$, $H_1(\epsilon)$, may be characterized by a non-trivial representation of $\mathcal{G}_0$. Under the external symmetry perturbation, the Hamiltonian becomes, 
\begin{eqnarray}
H_0 \rightarrow H(\epsilon) = H_0(\epsilon) + H_1(\epsilon), \quad H_0(\epsilon=0)=H_0,
\end{eqnarray}  
and $H(\epsilon \neq 0)$ enjoys a smaller symmetry group, $\mathcal{G}_1$. Note that all coupling constants in $H_0$ become functions of $\epsilon$, manifested by $H_0(\epsilon)$, and both $H_0(\epsilon)$ and $H_1(\epsilon)$ operators are trivial under $\mathcal{G}_1$ though $H_1(\epsilon \neq0)$ is non-trivial under $\mathcal{G}_0$. 
We focus on a perturbative regime, $|\epsilon| \ll 1$, which may be justified by the presence of a Mott gap in this work. 

We identify all possible spin interactions and determine their coefficients via symmetry analysis, DFT, and microscopic calculations as outlined below.
\begin{enumerate}
\item Identify all possible emergent spin interactions under external perturbations $(\epsilon)$ through symmetry analysis, denoted as $H_1(\epsilon)$.
\item Perform DFT calculation for a given material under the symmetry-breaking perturbation.
\item Construct the tight-binding model incorporating symmetry-breaking effects induced by external perturbations $(\epsilon)$.
\item Perform microscopic calculations to determine the coefficients of $H_0(\epsilon)$ and $H_1(\epsilon)$, and construct the generalized spin model, $H(\epsilon)$.
\end{enumerate}

 Below, we apply the general strategy to $\alpha$-RuCl$_3$
 and construct a generalized spin model under strain effects. With the estimated energy scales induced by $H_1(\epsilon)$, we propose and discuss the possible engineering of ground states via strain.\\

\begin{figure*}[tb]
\centering
\includegraphics[width=0.7\textwidth]{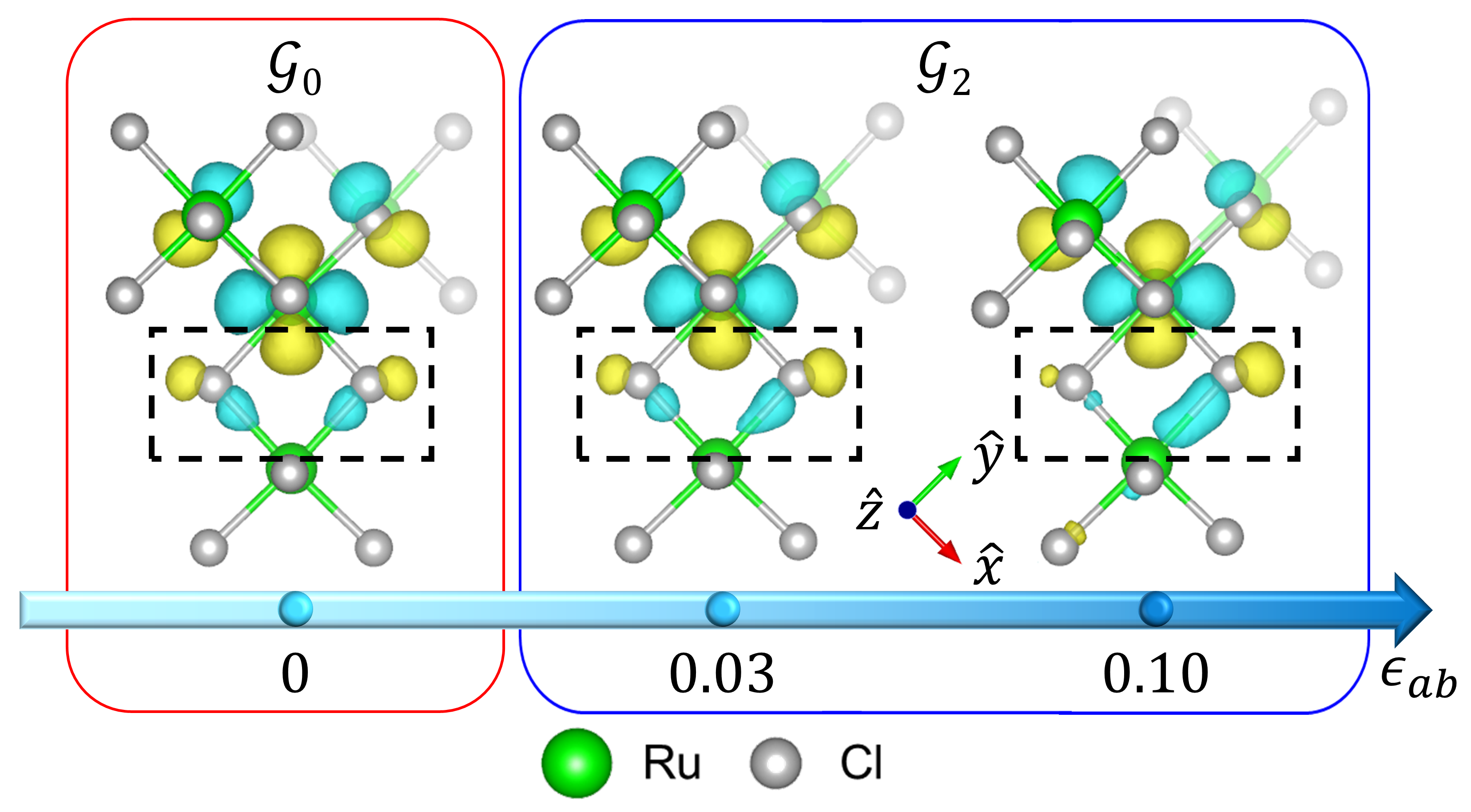}
\caption{ The visualization of  $d_{xy}$ Wannier function as a function of $D^{\alpha}_a$ strain strength. Namely, the system evolves from $\mathcal{G}_0$ to $\mathcal{G}_2$. The yellow and cyan color represents the positive and negative isosurfaces of Wannier functions, respectively. The strain-induced $C_{2}$ symmetry breaking (along the $z$-bond) results in the asymmetric spatial distribution  which is particularly pronounced in the boxed region. It also leads to  asymmetric hopping integrals.}
\label{fig2}
\end{figure*}

\noindent\textbf{Model and symmetry.} Let us consider a generic spin-$1/2$ model without strain effects on the honeycomb lattice, which is commonly referred to as the $KJ\Gamma \Gamma'$ model \cite{Kitaev2005.10,Rau2014.2}. 
\begin{align}
H_{0}=&\sum_{\langle j,k \rangle_{\gamma}} \large[KS_j^{\gamma}S_k^{\gamma}+J\mathbf{S}_j\cdot\mathbf{S}_k+\Gamma(S^{\alpha}_jS^{\beta}_k+S^{\beta}_jS^{\alpha}_k)\nonumber \\
&+\Gamma'(S^{\alpha}_jS^{\gamma}_k+S^{\gamma}_jS^{\alpha}_k+S^{\beta}_jS^{\gamma}_k+S^{\gamma}_jS^{\beta}_k)\large],
\end{align}
where $\braket{i,j}_{\gamma}$ are for the nearest-neighbor bonds with a component $\gamma \in \{x, y, z\}$ and $S^{\gamma}_{j,k}$ are $\gamma$ component spin operator at sites $j,k$.

We apply external strain and stress, assuming the generalized Hooke's law, $\sigma_{ij}=c_{ijkl}\epsilon_{kl}$ where $\boldsymbol{\sigma} (\epsilon)$ is Cauchy stress tensor  (infinitesimal strain tensor) and $\mathbf{c}$ is stiffness tensor. 
Their specific forms are as follows:
\begin{align}
\sigma_{ij}=\frac{d F_i}{dS^{(j)}},\quad \epsilon_{i,j}=\frac{1}{2}\left(\frac{\partial u_i}{\partial x_j}+\frac{\partial u_j}{\partial x_i}\right),
\end{align}
where $F_i$ is an $i$-component force, $S^{(n)}$ is surface with normal vector $n$, and $\mathbf{u}$ is the displacement field. 
Both normal ($\epsilon_{n,n}$) and sheer strain ($\epsilon_{i,j}, i\neq j$) are illustrated in Fig.~\ref{fig1}a.
Introducing homogeneous strain ($\epsilon$) to the system, the coupling constants become a function of the strain, for example $J \rightarrow J(\epsilon)$ and $K \rightarrow K(\epsilon)$.

In the highest-symmetry structure of $\alpha$-RuCl$_3$, the system enjoys the symmetry group, $\mathcal{G}_0$, which consists of the spatial inversion $(\mathcal{P})$ and the time-reversal $(\mathcal{T})$ symmetry in combination with a dihedral group ($D_3$).
The $D_3$ consists of three-fold rotations $(C_3)$ around the out-of-plane direction and two-fold rotations $(C_2)$ along the bond directions.
Under strains, $\mathcal{G}_0$ can be either maintained or broken down to lower symmetry groups. We uncover that only two groups, $\mathcal{G}_1$ and $\mathcal{G}_2$, are allowed, determined by the direction of $\epsilon$:
The six independent strains can be categorized into three groups, depending on the symmetry-breaking patterns.
\small{
\begin{align*}
&\text{$\mathcal{G}_0$ : (point group: $D_3\otimes \mathcal{P}$, space group: $P\bar{3}1m$)}=\{\epsilon_{a},\epsilon_{c}\},\\ 
&\text{$\mathcal{G}_1$ : (point group: $C_{2}\otimes \mathcal{P}$, space group: $C2/m$)}=\{D^{\alpha}_{b},D^{\beta}_{b}\},\\ 
&\text{$\mathcal{G}_2$ :  (point group: $\mathcal{P}$, space group: $P\bar{1}$)}=\{D^{\alpha}_{a},D^{\beta}_{a}\},
\end{align*}}
where the strains ($\epsilon_{a,c},D^{\alpha,\beta}_{a,b}$) follow the notation of Table~\ref{table1}. The most symmetric combination of the perturbations (e.g., $\epsilon_a$ and $\epsilon_c$) does not break any symmetry, preserving the original symmetry group, $\mathcal{G}_0$. As presented in Fig.~\ref{fig1}a (see the red-colored boxes), the $C_3$ and $C_2$ symmetry are well-maintained and the interactions are enforced to be equal for each bond direction. The less symmetric group, $\mathcal{G}_1$, is realized by applying strains with special combinations such as  $D^{\alpha}_b$ and $D^{\beta}_b$, which preserve the $C_2$ rotational symmetry only along the $z$-bond but break the $C_3$ symmetry of $\mathcal{G}_0$, as shown in Fig.~\ref{fig1}a (the green-colored boxes). Consequently, the $z$-bond interaction becomes different whereas the $x$- and $y$-bond interactions remain equal. The most generic combination of $D^{\alpha}_a$ and $D^{\beta}_a$ only enjoys the least symmetric group, $\mathcal{G}_2$, where both $C_3$ and $C_2$ rotational symmetry are broken, as shown in Fig.~\ref{fig1}a (blue-colored). In this case, all interactions are no longer equivalent.

All irreducible representations of the nearest two spins and the strains are determined, as shown in Table~\ref{table1}. We construct the additional strain-induced Hamiltonian ($H_1$) to the $KJ\Gamma\Gamma'$ model ($H_0$), referred to as the $\epsilon$-$KJ\Gamma\Gamma'$ model ($H_0+H_1$).
\small{
\begin{align}
H_{1}(\epsilon)=&+\mathbf{K}_{(\mathcal{E})}\cdot \overrightarrow{\mathcal{S}}^{K_{(\mathcal{E})}}+\mathbf{J}_{(\mathcal{E},1)}\cdot \overrightarrow{\mathcal{S}}^{J_{(\mathcal{E},1)}}+\mathbf{J}_{(\mathcal{E},2)}\cdot \overrightarrow{\mathcal{S}}^{J_{(\mathcal{E},2)}} \nonumber\\
&+J_{(\mathcal{A}_2)} \mathcal{S}^{J_{(\mathcal{A}_2)}}+\mathbf{\Gamma}_{(\mathcal{E})}\cdot \overrightarrow{\mathcal{S}}^{\Gamma_{(\mathcal{E})}}+\mathbf{\Gamma'}_{(\mathcal{E},1)}\cdot \overrightarrow{\mathcal{S}}^{\Gamma'_{(\mathcal{E},1)}} \nonumber \\
&+\mathbf{\Gamma'}_{(\mathcal{E},2)}\cdot \overrightarrow{\mathcal{S}}^{\Gamma'_{(\mathcal{E},2)}}+\Gamma'_{(\mathcal{A}_2)} \mathcal{S}^{\Gamma'_{(\mathcal{A}_2)}}, 
\end{align}} 
\normalsize{where $\mathcal{S}^{A_{(\mu)}}$} denotes emergent additional spin interactions constructed from linear combinations of $A$-type spin interactions corresponding to the $\mu$-irreducible representation of the $D_3$ group, and $A_{(\mu)}$ represents their corresponding coupling constants. For example, linear combinations of $K$-type spin interactions can be expressed,
\begin{align*}
c_1\sum_{\braket{j,k}_{x}} S^{x}_{j}S^{x}_{k}+c_2\sum_{\braket{j,k}_{y}}S^{y}_{j}S^{y}_{k}+c_3\sum_{\braket{j,k}_{z}}S^{z}_{j}S^{z}_{k}.   
\end{align*}
There are three independent linear combinations, represented by the real parameters $c_1$, $c_2$, and $c_3$. The case $c_1 = c_2 = c_3$ corresponds to the well-known isotropic Kitaev interaction, while the remaining two independent combinations correspond to $(\overrightarrow{\mathcal{S}}^{K_{(\mathcal{E})}})_{a}$ and $(\overrightarrow{\mathcal{S}}^{K_{(\mathcal{E})}})_{b}$ in Table~\ref{table1}, and they span the $\mathcal{E}$-representation. See the Supplemental Information for the explicit matrix representation of $\epsilon$-$KJ\Gamma\Gamma'$ model ($H_0+H_1$).

The coupling constants are expressed as polynomial functions of the strain ($\epsilon$), which are classified into three types according to the leading order in strain. See the Supplemental Information for a more detailed and explicit analysis.
\small{
\begin{align*}
&\text{Type \Romannum{1}: } \{K,J,\Gamma,\Gamma'\} \\ & \quad \quad \quad \quad =L(C_0)+O(\epsilon),\\ 
&\text{Type \Romannum{2}: }\{\mathbf{K}_{(\mathcal{E})},\mathbf{J}_{(\mathcal{E},1)},\mathbf{J}_{(\mathcal{E},2)},\mathbf{\Gamma}_{(\mathcal{E})},\mathbf{\Gamma}'_{(\mathcal{E},1)},\mathbf{\Gamma}'_{(\mathcal{E},2)}\}\\
&\quad \quad \quad \quad =L(\mathbf{D}^{\alpha},\mathbf{D}^{\beta})+O(\epsilon^2), \label{types}\\ 
&\text{Type \Romannum{3}: }\{ J_{(\mathcal{A}_2)},\Gamma'_{(\mathcal{A}_2)}\} \\ & \quad \quad \quad \quad =L(\mathbf{D}^{\alpha}\times \mathbf{D}^{\beta})+O(\epsilon^3),
\end{align*}
}
\normalsize{where} $C_0$ is a real constant, and $L(a_1, a_2,\cdots)$  denotes a linear combination of $a_1, a_2, \cdots$. 
 Each type falls into different irreducible representations of $D_3$ whose leading order of strain is $O(\epsilon^{n-1})$ for Type $n$. Type \Romannum{1} is trivial under all elements of $D_3$ $(\mathcal{A}_1)$, while Type \Romannum{3} is even under $C_3$ rotations and odd under $C_2$ rotations $(\mathcal{A}_2)$. Lastly, Type \Romannum{2} behaves like  $(\hat{a},\hat{b})$ vector under $D_3$ $(\mathcal{E})$.

We find the conditions of zero coupling constants by using the symmetries. 
First, under the  $\mathcal{G}_0$ symmetry, all Type \Romannum{2} and Type \Romannum{3} coupling constants are identically zero because they are non-trivial under $\mathcal{G}_0$. Second, under the $\mathcal{G}_1$ symmetry, $a$-components of vectors in Type \Romannum{2} and all Type \Romannum{3} coupling constants vanish. The symmetry-protected zero coupling constants allow us to cross-check the consistency of our symmetry analysis with the microscopic calculations, while the leading-order strain dependence enables a separate comparison with DFT results discussed below.\\

\begin{table*}[tb]
\begin{center}
\centering 
\begin{tabular}{>{\centering}p{1.5cm}|>{\centering\arraybackslash} p{14.0cm}}
\hline 
\hline 
& Interaction Coefficients  \\ 
\hline \hline
$K$ & $\frac{1}{6}\left(2C_{3}[t^{z}_{KJ}]+2C^2_{3}[t^{z}_{KJ}]+2t^{z}_{KJ}-C_{3}[t^{x}_{KJ}]-C^2_{3}[t^{x}_{KJ}]-t^{x}_{KJ}-C_{3}[t^{y}_{KJ}]-t^{y}_{KJ}-C^2_{3}[t^{y}_{KJ}]\right)$\\
$J$  & $\frac{1}{6}\left(C_{3}[t^{x}_{KJ}]+C^2_{3}[t^{x}_{KJ}]+t^{x}_{KJ}+C_{3}[t^{y}_{KJ}]+t^{y}_{KJ}+C^2_{3}[t^{y}_{KJ}]\right)$ \\
$\Gamma$ & $\frac{1}{3}\left(t^{z}_{\Gamma\Gamma'}+C_{3}[t^{z}_{\Gamma\Gamma'}]+C^2_{3}[t^{z}_{\Gamma\Gamma'}]\right)$\\
$\Gamma'$&$\frac{1}{6}\left(t^{x}_{\Gamma\Gamma'}+C_{3}[t^{x}_{\Gamma\Gamma'}]+C^2_{3}[t^{x}_{\Gamma\Gamma'}]+C_{2x}[t^{x}_{\Gamma\Gamma'}]+C_{2x}C_{3}[t^{x}_{\Gamma\Gamma'}]+C_{2x}C^2_{3}[t^{x}_{\Gamma\Gamma'}]\right)$\\[0.2cm]
\hline 
$(\mathbf{K}_{(\mathcal{E})})_b$ & $\frac{1}{\sqrt{6}}\left(C_{3}[t^{z}_{KJ}]+C^2_{3}[t^{z}_{KJ}]-2t^{z}_{KJ}\right)$\\
$(\mathbf{J}_{(\mathcal{E},1)})_b$ & $\frac{1}{2\sqrt{3}}\left(-\sqrt{3}C_{3}[t^{x}_{KJ}]+\sqrt{3}C^2_{3}[t^{x}_{KJ}]+\sqrt{3}C_{3}[t^{y}_{KJ}]-\sqrt{3}C^2_{3}[t^{y}_{KJ}]\right)$\\
$(\mathbf{J}_{(\mathcal{E},2)})_b$ & $\frac{1}{2\sqrt{3}}\left( C_{3}[t^{x}_{KJ}]+C^2_{3}[t^{x}_{KJ}]-2t^{x}_{KJ}+C_{3}[t^{y}_{KJ}]-2t^{y}_{KJ}+C^2_{3}[t^{y}_{KJ}]\right)$\\
$(\boldsymbol{\Gamma}_{(\mathcal{E})})_b$ & $\frac{1}{\sqrt{3}}\left( -2t^{z}_{\Gamma\Gamma'}+C_{3}[t^{z}_{\Gamma\Gamma'}]+C^2_{3}[t^{z}_{\Gamma\Gamma'}]\right)$\\
$(\boldsymbol{\Gamma}'_{(\mathcal{E},1)})_b$ & $\frac{1}{\sqrt{6}}\left( -\sqrt{3}t^{x}_{\Gamma\Gamma'}+\sqrt{3}C_{3}[t^{x}_{\Gamma\Gamma'}]+\sqrt{3}C_{2x}[t^{x}_{\Gamma\Gamma'}]-\sqrt{3}C_{2z}[t^{x}_{\Gamma\Gamma'}]\right)$\\
$(\boldsymbol{\Gamma}'_{(\mathcal{E},2)})_b$ & $\frac{1}{\sqrt{6}}\left( t^{x}_{\Gamma\Gamma'}+C_{3}[t^{x}_{\Gamma\Gamma'}]-2C^2_{3}[t^{x}_{\Gamma\Gamma'}]+C_{2x}[t^{x}_{\Gamma\Gamma'}]-2C_{2y}[t^{x}_{\Gamma\Gamma'}]+C_{2z}[t^{x}_{\Gamma\Gamma'}]\right)$\\[0.2cm]
\hline 
$(\mathbf{K}_{(\mathcal{E})})_a$ & $\frac{1}{\sqrt{2}}C_{3}[t^{z}_{KJ}]-\frac{1}{\sqrt{2}}C^2_{3}[t^{z}_{KJ}]$\\
$(\mathbf{J}_{(\mathcal{E},1)})_a$ & $\frac{1}{2\sqrt{3}}\left(C_{3}[t^{x}_{KJ}]+C^2_{3}[t^{x}_{KJ}]-2t^{x}_{KJ}-C_{3}[t^{y}_{KJ}]+2t^{y}_{KJ}-C^2_{3}[t^{y}_{KJ}]\right)$\\
$(\mathbf{J}_{(\mathcal{E},2)})_a$ & $\frac{1}{2\sqrt{3}}\left(\sqrt{3}C_{3}[t^{x}_{KJ}]-\sqrt{3}C^2_{3}[t^{x}_{KJ}]+\sqrt{3}C_{3}[t^{y}_{KJ}]-\sqrt{3}C^2_{3}[t^{y}_{KJ}]\right)$\\
$(\boldsymbol{\Gamma}_{(\mathcal{E})})_a$ & $C_{3}[t^{z}_{\Gamma\Gamma'}]-C^2_{3}[t^{z}_{\Gamma\Gamma'}]$\\
$(\boldsymbol{\Gamma}'_{(\mathcal{E},1)})_a$ & $\frac{1}{\sqrt{6}}\left(t^{x}_{\Gamma\Gamma'}+C_{3}[t^{x}_{\Gamma\Gamma'}]-2C^2_{3}[t^{x}_{\Gamma\Gamma'}]-C_{2x}[t^{x}_{\Gamma\Gamma'}]+2C_{2y}[t^{x}_{\Gamma\Gamma'}]-C_{2z}[t^{x}_{\Gamma\Gamma'}] \right)$\\
$(\boldsymbol{\Gamma}'_{(\mathcal{E},2)})_a$ & $\frac{1}{\sqrt{6}}\left(\sqrt{3}t^{x}_{\Gamma\Gamma'}-\sqrt{3}C_{3}[t^{x}_{\Gamma\Gamma'}]+\sqrt{3}C_{2x}[t^{x}_{\Gamma\Gamma'}]-\sqrt{3}C_{2z}[t^{x}_{\Gamma\Gamma'}] \right)$\\
\hline
$J_{(\mathcal{A}_2)}$ & $\frac{1}{6}(C_{3}[t^{x}_{KJ}]+C^2_{3}[t^{x}_{KJ}]+t^{x}_{KJ}-C_{3}[t^{y}_{KJ}]-t^{y}_{KJ}-C^2_{3}[t^{y}_{KJ}])$\\
$\Gamma'_{(\mathcal{A}_2)}$ & $\frac{1}{6}(t^{x}_{\Gamma\Gamma'}+C_{3}[t^{x}_{\Gamma\Gamma'}]+C^2_{3}[t^{x}_{\Gamma\Gamma'}]-C_{2x}t^{x}_{\Gamma\Gamma'}-C_{2x}C_{3}[t^{x}_{\Gamma\Gamma'}]-C_{2x}C^2_{3}[t^{x}_{\Gamma\Gamma'}])$\\[0.2cm]
\hline
\hline
\end{tabular}
\caption{ Coefficients of $\epsilon$-$KJ\Gamma\Gamma'$ model ($H_0+H_1$) in the presence of strain expressed through hopping terms. $C_3[\cdot]$ is a three-fold rotation around the out-of-plane direction and $C_{2\alpha}[\cdot]$ is a two-fold rotation along the $\alpha$-bond directions.
}
%
 \label{table2}
\end{center}
\end{table*}

\noindent \textbf{DFT calculations}.
While the electronic structure of RuCl$_3$ has been extensively investigated for both bulk and monolayer \cite{kim_kitaev_2015, Kim2016.4, sarikurt_electronic_2018, iyikanat_tuning_2018, vatansever_strain_2019, liu_contrasting_2023, samanta_electronic_2024}, the study of strain effects has been limited to the case of normal ones \cite{iyikanat_tuning_2018, vatansever_strain_2019}. As a result, the systematic investigation of generic strain situations and the corresponding  $j_\text{eff}=1/2$ electronic structure have been largely unexplored. Fig.~\ref{fig1}b shows the calculated band structure of the pristine (no strain) RuCl$_3$ monolayer in its zigzag antiferromagnetic ground state. It is in good agreement with the previous study \cite{kim_kitaev_2015, iyikanat_tuning_2018, vatansever_strain_2019}. Due to the sizable SOC and on-site Coulomb interaction, the near-Fermi-level states are well characterized by $j_\text{eff}=1/2$ and $j_\text{eff}=3/2$ with sizable `Mott' gap.

Figure~\ref{fig1}c shows the $j_\text{eff}$-projected band dispersion under 3\% $\epsilon_{ac}$ strain,  corresponding to $D^{\beta}_b$. The result is not much different from Fig.~\ref{fig1}b. 
Figure~\ref{fig1}d shows the result of $\mathcal{G}_2$ group, namely under $-$3\% $\epsilon_{ab}$ ($D^{\alpha}_a$) strain, where the different high symmetry points are selected due to monoclinic deformation. 
The $j_\text{eff}=1/2$ character of upper Hubbard bands is well maintained while some  noticeable differences are clearly identified (e.g., the lifted degeneracy at X-point, corresponding to S-point in Figs.~\ref{fig1}b and \ref{fig1}c). 
The main features are  also well preserved in the case of $D^{\beta}_a$
strain; see Supplementary Information Fig.~S2.

%
%
%
%
%


To examine the microscopic change in response to symmetry-breaking strains, we visualize Ru-$t_\text{2g}$ Wannier orbitals as a function of $\epsilon_{ab}$ strain, see Fig.~\ref{fig2}. For the $z$-bond, neighboring Ru atoms lie within the $xy$ plane, whereas for the $x$- and $y$-bonds, the neighboring Ru atoms locate outside this plane. Under zero strains ($\mathcal{G}_0$), $d_{xy}$,  $d_{yz}$, and $d_{zx}$ orbitals should be symmetric with respect to $C_2$ rotation about the $z$-, $x$-, and $y$-bond directions, respectively. It is indeed well observed in our plot of $d_{xy}$ in Fig.~\ref{fig2} (the left most figure). As the strength of $\epsilon_{ab}$ strain increases, the inequivalent spatial distribution becomes clear; see, e.g., the dashed box region. Namely, under $D^{\alpha}_a$ strain, the system belonging to $\mathcal{G}_2$ group lacks $C_2$ symmetry, which is responsible for asymmetric electron distribution and hopping integral.\\

\noindent\textbf{Microscopics.}
We derive the effective Hamiltonian in the presence of strain effects from a microscopic perspective and compare it with the symmetry analysis presented in the Model and Symmetry section. Our DFT results show that, even under strain, the $j_{\text{eff}}=1/2$ and $j_{\text{eff}}=3/2$ states remain well defined. This allows us to perform a perturbative calculation within the $j_{\text{eff}}=1/2$ subspace by incorporating symmetry-breaking effects due to strain. 

First, we consider an atomic Hamiltonian of Kanamori type to carry out the strong coupling expansion for holes for each Ru$^{3+}$ ion \cite{Rau2014.2,Kanamori}, 
\begin{align*}
&H_{h}=\frac{(U-3J_{\text{H}})}{2}(N-1)^2-2J_{\text{H}}S^2-\frac{J_{\text{H}}}{2}L^2+\lambda_{\text{SO}} \mathbf{S}\cdot \mathbf{L} 
\end{align*}
where $U$, $J_\text{H}$, and $\lambda_\text{SO}$ denote the on-site Coulomb repulsion, Hund’s coupling, and spin-orbit interaction, respectively, and $N$, $S$, and $L$ are the total hole number, spin, and orbital angular momentum operators for each hole.

Next, we consider the kinetic Hamiltonian, which reflects the symmetry breaking induced by strain, as a perturbation.
\begin{align}
H_{\text{kin}}=\sum_{\langle i,j \rangle_{\gamma}}\left(\overrightarrow{a}^{\dagger}_{i}\overleftrightarrow{t}_{ij}^{\gamma}\overrightarrow{a}_{j}+\overrightarrow{a}^{\dagger}_{j}\overleftrightarrow{t}_{ji}^{\gamma}\overrightarrow{a}_{i}\right)
\end{align}
where  $\overrightarrow{a}^{\dagger}_{i}=(a^{\dagger}_{i,yz},a^{\dagger}_{i,zx},a^{\dagger}_{i,xy})$, $a^{\dagger}_{i,yz}=(a^{\dagger}_{i,yz,\uparrow},a^{\dagger}_{i,yz,\downarrow})$, and $a_{i,l,s}$ represent the creation and annihilation operators for $t_{\text{2g}}$ orbitals at site $i$ with the orbital $l$ and spin $s$.
The original group structure $\mathcal{G}_0$ allows symmetry-restricted kinetic matrices, 
\begin{widetext}
\begin{align}
\mathcal{G}_0&:\quad \overleftrightarrow{t}_{ij}^{x}=
\begin{pmatrix}
t_{xx}^{x} & t_{xy}^{x} &t_{xy}^{x} \\
t_{xy}^{x} & t_{yy}^{x} &t_{yz}^{x} \\
t_{xy}^{x} & t_{yz}^{x} &t_{yy}^{x} \\
\end{pmatrix}, \overleftrightarrow{t}_{ij}^{y}=
\begin{pmatrix}
t_{yy}^{x} & t_{xy}^{x} &t_{yz}^{x} \\
t_{xy}^{x} & t_{xx}^{x} &t_{xy}^{x} \\
t_{yz}^{x} & t_{xy}^{x} &t_{yy}^{x} \\
\end{pmatrix},
\overleftrightarrow{t}_{ij}^{z}=
\begin{pmatrix}
t_{yy}^{x} & t_{yz}^{x} &t_{xy}^{x} \\
t_{yz}^{x} & t_{yy}^{x} &t_{xy}^{x} \\
t_{xy}^{x} & t_{xy}^{x} &t_{xx}^{x} \\
\end{pmatrix} \label{g0}. 
\end{align}
\end{widetext}

The full $D_3$ symmetry allows only 4 degrees of freedom ($t_{xx}^{x}, t_{xy}^{x},  t_{yy}^{x},  t_{yz}^{x}$), which are precisely equivalent to the previous results in literature \cite{Rau2014.2}.

When strain effects reduce the original symmetry group to $\mathcal{G}_1$, the kinetic matrices become
\begin{widetext}
\begin{align}
\mathcal{G}_1&:\quad \overleftrightarrow{t}_{ij}^{x}=
\begin{pmatrix}
t_{xx}^{x} & t_{xy}^{x} &t_{xz}^{x} \\
t_{xy}^{x} & t_{yy}^{x} &t_{yz}^{x} \\
t_{xz}^{x} & t_{yz}^{x} &t_{yy}^{x} \\
\end{pmatrix}, \overleftrightarrow{t}_{ij}^{y}=
\begin{pmatrix}
t_{yy}^{x} & t_{xy}^{x} &t_{yz}^{x} \\
t_{xy}^{x} & t_{xx}^{x} &t_{xz}^{x} \\
t_{yz}^{x} & t_{xz}^{x} &t_{zz}^{x} \\
\end{pmatrix},
\overleftrightarrow{t}_{ij}^{z}=
\begin{pmatrix}
t_{xx}^{z} & t_{xy}^{z} &t_{xz}^{z} \\
t_{xy}^{z} & t_{xx}^{z} &t_{xz}^{z} \\
t_{xz}^{z} & t_{xz}^{z} &t_{zz}^{z} \\
\end{pmatrix}. \label{g1}
\end{align}
\end{widetext}
The lower symmetry group allows more coupling constants, giving 10 hopping degrees of freedom with  $C_{2z}$ symmetry. Note that this symmetry group has been reported in previous experiments in  $\alpha$-RuCl$_3$ \cite{Fletcher1967.1,Kubota2015.3}.

\begin{table*}[tb]
\begin{center}
\centering 
\begin{tabular}{>{\centering}p{1.5cm}|>{\centering}p{1.5cm}|>{\centering}p{1.5cm}|>{\centering}p{1.5cm}|>{\centering}p{1.5cm}|>{\centering}p{1.5cm}|>{\centering}p{1.5cm}|>{\centering}p{1.5cm}|>{\centering\arraybackslash} p{1.5cm}}
\hline 
\hline 
& Pristine&$\epsilon_a$&$\epsilon_c$&$D^{\alpha}_b$& $D^{\beta}_b$ &$D^{\alpha}_a$&$D^{\beta}_a$&$D^{\alpha}_a+D^{\beta}_b$\\ 
\hline \hline
Group& \multicolumn{3}{c|}{$\mathcal{G}_0=P\bar{3}1m$}&\multicolumn{2}{c|}{$\mathcal{G}_1=C2/m$}&\multicolumn{3}{c}{$\mathcal{G}_2=P\bar{1}$}\\ 
\hline \hline
$K$ &$-6.595$ &$-6.469$&$-4.726$&
$-6.259$&$-6.741$&$-6.297$&$-6.731$&$-6.441$\\
$J$ &$-1.014$ &$-0.377$&$-1.395$&
$-1.207$&$-0.997$&$-1.181$&$-0.994$&$-1.176$\\
$\Gamma$ &$3.906$ &$2.178$ &$4.465$ &
$4.070$ &$3.683$ &$4.059$ &$3.678$&$3.791$\\
$\Gamma'$ &$-0.857$ &$-0.589$ &$-0.859$ &
$-0.853$ &$-0.906$ &$-0.851$ &$-0.902$&$-0.905$\\[0.1cm]
\hline
$(\mathbf{K}_{(\mathcal{E})})_b$ &
\multirow{6}{*}{0} &\multirow{6}{*}{0} &\multirow{6}{*}{0} &$1.128$ &$5.164$ &$0.201$ &$-0.155$&$5.335$\\
$(\mathbf{J}_{(\mathcal{E},1)})_b$ &
&&&$-1.767$ &$0.443$ &$0.275$ &$0.029$&$0.678$\\
$(\mathbf{J}_{(\mathcal{E},2)})_b$ &
&&&$-1.263$ &$0.590$ &$0.237$ &$-0.008$&$0.864$\\
$(\boldsymbol{\Gamma}_{(\mathcal{E})})_b$ &
&&&$-5.353$ &$0.695$ &$0.502$ &$-0.206$&$1.160$\\
$(\boldsymbol{\Gamma}'_{(\mathcal{E},1)})_b$ &
&&&$1.036$ &$-0.595$ &$0.055$ &$-0.104$&$-0.513$\\
$(\boldsymbol{\Gamma}'_{(\mathcal{E},2)})_b$ &
&&&$-0.539$ &$1.580$ &$0.043$ &$-0.125$&$1.609$\\[0.1cm]
\hline
$(\mathbf{K}_{(\mathcal{E})})_a $&
\multirow{6}{*}{0} &\multirow{6}{*}{0} &\multirow{6}{*}{0} &\multirow{6}{*}{0} &\multirow{6}{*}{0} &$1.324$ &$5.031$&$0.792$\\
$(\mathbf{J}_{(\mathcal{E},1)})_a$ &
&&& & &$-1.489$ &$0.473$&$-1.634$\\
$(\mathbf{J}_{(\mathcal{E},2)})_a$ &
&&& & &$-1.029$ &$0.583$&$-1.043$\\
$(\boldsymbol{\Gamma}_{(\mathcal{E})})_a$ &
&&& & &$-4.853$ &$0.488$&$-4.363$\\
$(\boldsymbol{\Gamma}'_{(\mathcal{E},1)})_a$ &
&&& & &$1.089$ &$-0.708$&$1.138$\\
$(\boldsymbol{\Gamma}'_{(\mathcal{E},2)})_a$ &
&&& & &$-0.499$&$1.455$&$-0.451$\\
\hline
$J_{(\mathcal{A}_2)}$ &
\multirow{2}{*}{0}&\multirow{2}{*}{0}&\multirow{2}{*}{0}&\multirow{2}{*}{0}&\multirow{2}{*}{0}  &$0.010$&$0.003$&$0.076$\\
$\Gamma'_{(\mathcal{A}_2)}$ &
&&& & &$-0.007$&$-0.004$&$-0.047$\\[0.1cm]
\hline
\hline\end{tabular}
\caption{DFT results for the coefficients of $\epsilon$-$KJ\Gamma\Gamma'$ model ($H_0+H_1$)  under each $3\%$ strain. The values are given in \si{\milli\electronvolt}.}
 \label{table3}
\end{center}
\end{table*}

For generic strain effects, the lowest symmetry group  $\mathcal{G}_2$ gives 18 degrees of freedom, representing the most general form allowed under homogeneous strain.
\begin{widetext}
\begin{align}
\mathcal{G}_2&:\quad \overleftrightarrow{t}_{ij}^{x}=
\begin{pmatrix}
t_{xx}^{x} & t_{xy}^{x} &t_{xz}^{x} \\
t_{xy}^{x} & t_{yy}^{x} &t_{yz}^{x} \\
t_{xz}^{x} & t_{yz}^{x} &t_{zz}^{x} \\
\end{pmatrix}, \overleftrightarrow{t}_{ij}^{y}=
\begin{pmatrix}
t_{xx}^{y} & t_{xy}^{y} &t_{xz}^{y} \\
t_{xy}^{y} & t_{yy}^{y} &t_{yz}^{y} \\
t_{xz}^{y} & t_{yz}^{y} &t_{zz}^{y} \\
\end{pmatrix},
\overleftrightarrow{t}_{ij}^{z}=
\begin{pmatrix}
t_{xx}^{z} & t_{xy}^{z} &t_{xz}^{z} \\
t_{xy}^{z} & t_{yy}^{z} &t_{yz}^{z} \\
t_{xz}^{z} & t_{yz}^{z} &t_{zz}^{z} \\
\end{pmatrix}. \label{g2}
\end{align}
\end{widetext}
Note that the hopping matrices are real and symmetric because of time-reversal $(\mathcal{T})$ and spatial inversion $(\mathcal{P})$ symmetries. 

In the limit of $U,J_\text{H} \gg \lambda_\text{SO} \gg |t_{ab}^c|$, we perform standard perturbative calculations to derive the additional spin Hamiltonian ($H_1$) and determine the coefficients of the $\epsilon$-$KJ\Gamma\Gamma'$ model ($H_0+H_1$).
The explicit forms of the operators and the coupling constants are summarized in Table~\ref{table1} and Table~\ref{table2} with the  notation ($t^{\alpha}_{KJ},t^{\alpha}_{\Gamma \Gamma'}$), 
\begin{widetext}
\begin{align}
t^{\alpha}_{KJ}=&\frac{4}{27(U-J_{\text{H}})}\Big(-(t^{\alpha}_{zz})^{2}+2(t^{\alpha}_{yy})^{2}+9(t^{\alpha}_{xy})^{2}+2(t^{\alpha}_{xx})^{2}-5t^{\alpha}_{xx}t^{\alpha}_{yy}+t^{\alpha}_{zz}(t^{\alpha}_{xx}+t^{\alpha}_{yy})\Big) \nonumber \\
&+\frac{4}{27(U-3J_{\text{H}})}\Big(3(t^{\alpha}_{zz})^{2}-9(t^{\alpha}_{xy})^{2}+9t^{\alpha}_{xx}t^{\alpha}_{yy}+3t^{\alpha}_{zz}(t^{\alpha}_{xx}+t^{\alpha}_{yy})\Big) +\frac{4(t^{\alpha}_{zz}+t^{\alpha}_{xx}+t^{\alpha}_{yy})^2}{27(U+2J_{\text{H}})}, \nonumber \\
t^{\alpha}_{\Gamma\Gamma'}=&\frac{8J_{\text{H}}\Big(t^{\alpha}_{xy}(t^{\alpha}_{xx}+t^{\alpha}_{yy}-2t^{\alpha}_{zz})+3t^{\alpha}_{xz}t^{\alpha}_{yz}\Big)}{9(U-J_{\text{H}})(U-3J_{\text{H}})}. \nonumber
\end{align}
\end{widetext}
where the $t^{\alpha}_{KJ}$ terms are related to a generalization of the $K,J$-type interactions, while the $t^{\alpha}_{\Gamma,\Gamma'}$ terms correspond to a generalized form of the $\Gamma,\Gamma'$-type interactions.
It should be emphasized that the rotation ($C_3,C_{2\alpha}$) in Table~\ref{table2} must be applied to the hopping terms 
($t^{c}_{ab}$), rather than directly to the labels 
$\alpha$ appearing in $t^{\alpha}_{KJ}$ and $t^{\alpha}_{\Gamma,\Gamma'}$. For example, 
\begin{align*}
C_{3}[t^{x}_{\Gamma\Gamma'}]=\frac{8J_{H}\Big(t^{y}_{yz}(t^{y}_{yy}+t^{y}_{zz}-2t^{y}_{xx})+3t^{y}_{xy}t^{y}_{zx}\Big)}{9(U-J_{H})(U-3J_{H})}\neq t^{y}_{\Gamma\Gamma'}.   
\end{align*}

A straightforward way to validate our calculations is to check whether our model correctly reproduces the $KJ\Gamma\Gamma'$ model. The original Kitaev interaction ($K$) is modified to 
\begin{eqnarray}
K(\epsilon) &=\frac{1}{6}\big(2C_{3}[t^{z}_{KJ}]+2C^2_{3}[t^{z}_{KJ}]+2t^{z}_{KJ}-C_{3}[t^{x}_{KJ}] \nonumber\\
&-C^2_{3}[t^{x}_{KJ}]-t^{x}_{KJ}-C_{3}[t^{y}_{KJ}]-t^{y}_{KJ}-C^2_{3}[t^{y}_{KJ}]\big). \nonumber
\end{eqnarray}
It is straightforward to check the original symmetry group, $\mathcal{G}_0$ in Eq.~\eqref{g0}, restores the original Kitaev interaction term \cite{Rau2014.2},  
\begin{eqnarray}
K(\epsilon \rightarrow 0) = t^{z}_{KJ}-t^{x}_{KJ}=K. \nonumber 
\end{eqnarray}
The other interactions ($J, \Gamma, \Gamma^{'}$) are also restored in the limit of $\epsilon \rightarrow 0$. 

%

Furthermore, substituting Eq.~\eqref{g0} ($\mathcal{G}_0$) into the coupling constants listed in Table~\ref{table2} forces all Type \Romannum{2} and Type \Romannum{3} coupling constants to vanish identically, while Eq.~\eqref{g1} ($\mathcal{G}_1$) eliminates the $a$-components of the Type \Romannum{2} vectors as well as all Type \Romannum{3} coupling constants.
This behavior is in agreement with the discussion of symmetry-protected zero coupling constants in the Model and Symmetry section, demonstrating consistency with our microscopic calculations.\\

\noindent\textbf{DFT estimation of spin interactions.} We calculate spin interactions in our $\epsilon $-$KJ\Gamma \Gamma'$ model based on the microscopic analysis presented above. The hopping parameters are estimated in the standard way of combining DFT and Wannier function techniques. The results of six different strains are summarized in Table~\ref{table3}. It is clearly noted that $\mathcal{G}_0$, $\mathcal{G}_1$, and $\mathcal{G}_2$ group are well distinct by having only the allowed interaction parameters while each one of them properly reflects the structural deformations. 

For the case of $\mathcal{G}_0$ group, the results are well represented by conventional $KJ\Gamma\Gamma'$ model with modified magnitudes. Interestingly, both in-plane  biaxial tensile strain and out-of-plane uniaxial compressive strain can induce a unique phase close to the ideal ferromagnetic Kitaev phase. In the $\mathcal{G}_1$ group under $D^{\alpha}_b$ or $D^{\beta}_b$ strain, $C_{3}$  symmetry is broken and such anisotropic interactions emerge as $(\mathbf{K}_{(\mathcal{E})})_b$, $(\mathbf{J}_{(\mathcal{E},1)})_b$, $(\mathbf{J}_{(\mathcal{E},2)})_b$, $(\boldsymbol{\Gamma}_{(\mathcal{E})})_b$, $(\boldsymbol{\Gamma}'_{(\mathcal{E},1)})_b$, and $(\boldsymbol{\Gamma}'_{(\mathcal{E},2)})_b$. Notably, under sufficiently large strain ($\sim 3 \%$), certain anisotropic interactions become comparable in strength to the isotropic ones; see, e.g., $\lvert (\boldsymbol{\Gamma}_{(\mathcal{E})})_b/\Gamma \rvert \sim 1.32$ for $D^{\alpha}_b$ and $\lvert (\mathbf{K}_{(\mathcal{E})})_b/K \rvert \sim 0.77$ for $D^{\beta}_b$ in Table~\ref{table3}. It highlights the limitation of $KJ\Gamma\Gamma'$ model in describing the strain effects.
The results of the $\mathcal{G}_2$ group clearly demonstrate that all interactions can have non-zero strengths. Meanwhile, the terms absent in our first-order perturbation exhibit relatively small values; see $(\mathbf{K}_{(\mathcal{E})})_b$, $(\mathbf{J}_{(\mathcal{E},1)})_b$, $(\mathbf{J}_{(\mathcal{E},2)})_b$, $(\boldsymbol{\Gamma}_{(\mathcal{E})})_b$, $(\boldsymbol{\Gamma}'_{(\mathcal{E},1)})_b$, and $(\boldsymbol{\Gamma}'_{(\mathcal{E},2)})_b$ in Table~\ref{table3}.

Together with the previous symmetry analysis in Eq.~\eqref{types}, we can make several observations for the case of $\mathcal{G}_2$ group. First, in Type \Romannum{1}, the deviation from the pristine case appears only in $O(\epsilon^2)$ and is therefore relatively small. Second, for Type \Romannum{2}, the coupling constants are predominantly governed by the linear order of strains under symmetry, and any deviation from linearity reflects the contributions from $O(\epsilon^2)$. Lastly, in the case of Type \Romannum{3}, the leading-order term is composed of only a specific combination of strain components $(\mathbf{D}^{\alpha}\times \mathbf{D}^{\beta})$, so the leading order of strain dependence differs depending on whether this combination vanishes or not. These analyses are further supported by the quantitative DFT results, as shown in Table~\ref{table3}, once again confirming the consistency of our findings.\\



%
%
\begin{figure}[tb]
\centering
\includegraphics[width=0.48\textwidth]{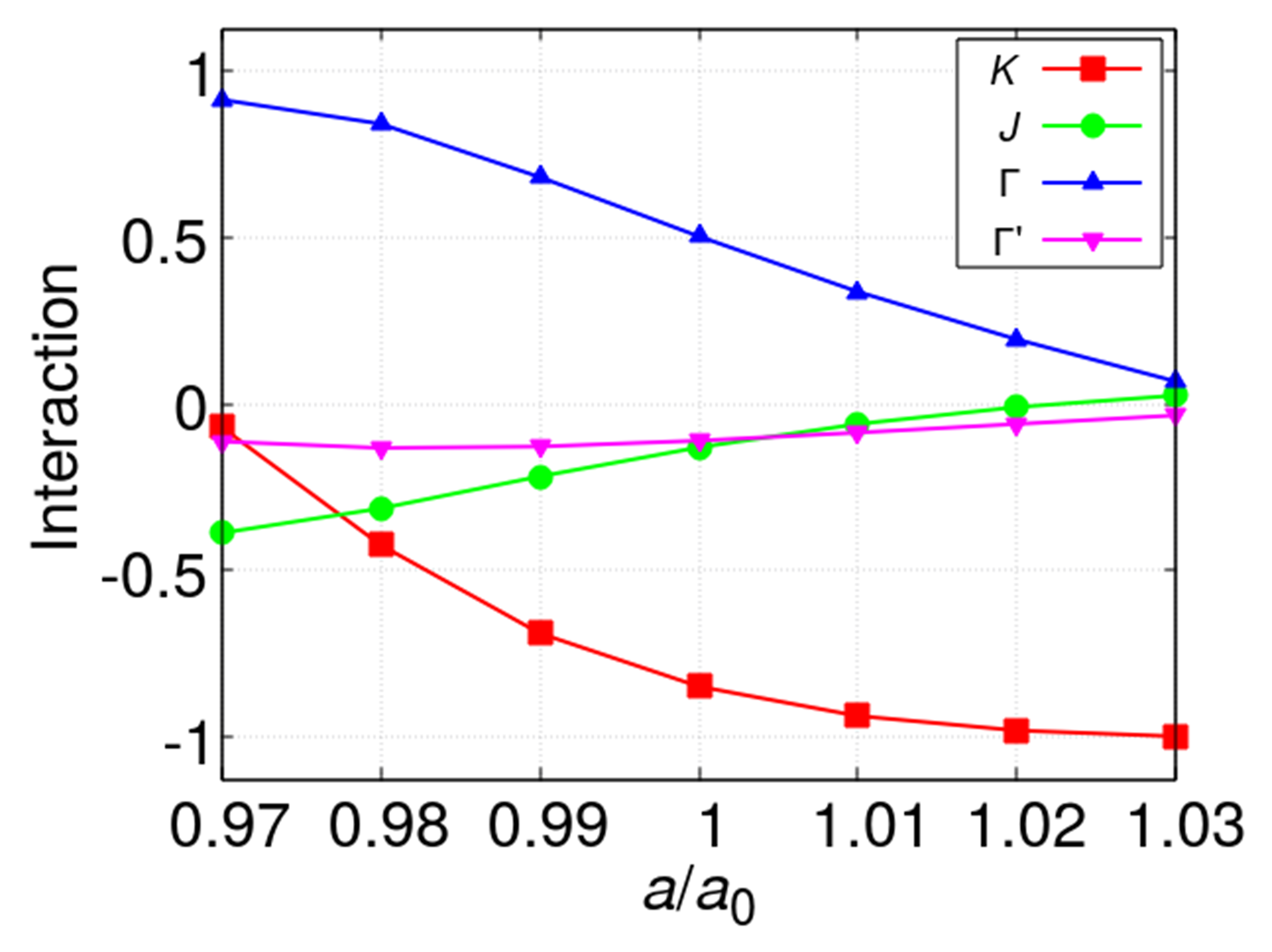}
\caption{The calculated spin interactions of $\alpha$-RuCl$_3$ monolayer under the in-plane biaxial strain with the fixed cell volume. Each interaction was normalized by $\sqrt {{J^2} + {K^2} + {\Gamma ^2} + \Gamma {'^2}} $. Red square, green circle, blue triangle, and magenta inverted triangle represent $K$, $J$, $\Gamma$, and $\Gamma'$, respectively.}
\label{fig3}
\end{figure}

\begin{table}[tb]
\centering 
\begin{tabular}{>{\centering}p{4cm}||>{\centering}p{1cm}|>{\centering}p{1cm}|>{\centering\arraybackslash} p{1cm}}
\hline 
\hline 
 Physical quantities & $\mathcal{T}$ & $\mathcal{P}$  & $\mathbb{D}_3$ \\ 
\hline \hline 
$h_c$$\, \propto (h_x + h_y + h_z)$  & odd & even & $\mathcal{A}_2$ \\
$h_c^3$ & odd & even & $\mathcal{A}_2$ \\
$h_c(h_a^2+h_b^2)$ & odd & even & $\mathcal{A}_2$ \\
$h_a(h_a^2-3h_b^2)$$\, \propto h_x h_y h_z$  & odd & even & $\mathcal{A}_2$ \\
\hline
$h_c\epsilon_{a}$ & odd & even & $\mathcal{A}_2$  \\
$h_c\epsilon_{c}$ & odd & even & $\mathcal{A}_2$  \\
$\mathbf{D}^{\alpha}\times(h_a,h_b) $ & odd & even & $\mathcal{A}_2$  \\
$ \mathbf{D}^{\beta}\times(h_a,h_b)$ & odd & even & $\mathcal{A}_2$  \\
\hline 
\hline 
\end{tabular} 
\caption{All possible irreducible representations that are odd for time-reversal, even for spacial inversion, and $\mathcal{A}_2$ irreducible representation.} 
\label{table4}
\end{table}

\begin{figure*}[tb]
\centering
\includegraphics[width=0.9\textwidth]{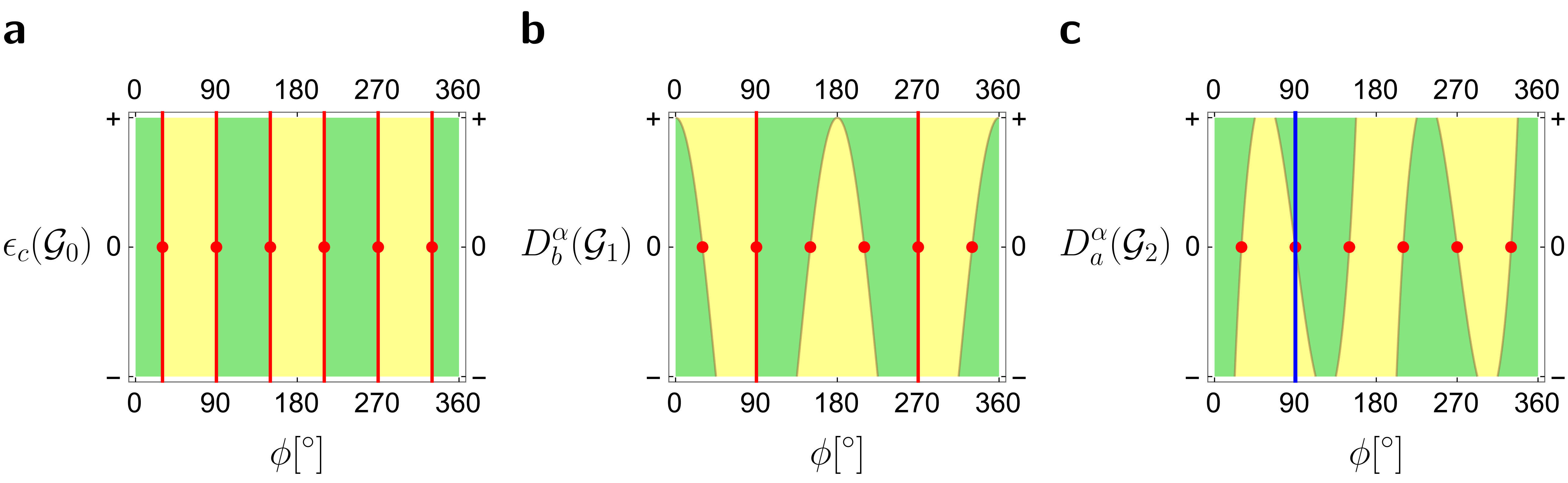}
\caption{Schematic strain-driven topological phase transitions. Figures \textbf{\textsf{a}}-\textbf{\textsf{c}} show the topological invariant $(\nu)$ for strain conditions $\epsilon_c$, $D^{\alpha}_b$, and $D^{\alpha}_a$, respectively, representing each different symmetry group when $k_{1\sim 8}=1$ in \eqref{eqnu} and the magnetic field is constrained in-plane ($\mathbf{h}=\cos{\phi}\hat{\mathbf{a}}+\sin{\phi}\hat{\mathbf{b}}$). 
The green (yellow) area represents the topological invariant of $1\;(-1)$, and the red dots and lines indicate topological phase transition points protected by symmetry. The blue line is the proposed setup for measuring the strain-driven topological phase transition.}
\label{fig4}
\end{figure*}

\noindent\textbf{Strain-driven quantum phase transitions.} Our result provides two important implications. First, the strain can induce a quantum phase transition  by tuning the strengths of the coupling constants only. Second, it suggests strain-induced quantum phase transitions between phases with different symmetry breaking patterns. For the case of $\mathcal{G}_0$ symmetry preserved, the conventional $KJ\Gamma \Gamma'$ model has been analyzed extensively in literature, and the phase diagrams with various magnetic structures such as zigzag antiferromagnetic order and ferromagnetic order have been suggested in terms of relative strengths of the coupling constants, ($K, J, \Gamma, \Gamma'$). \cite{Rau2014.2, rau_trigonal_2014} Then, it is obvious that our results of the $\epsilon$-$KJ\Gamma \Gamma'$ model with the normal strain, for example $\epsilon_a-\epsilon_c$, are directly applicable to the previous calculations, and we found that the ideal KQSL phase with ferromagnetic Kitaev interaction can be stabilized for a set of parameters. Figure~\ref{fig3} shows that tensile strains make $K$ interaction (red squares) dominant by markedly suppressing the other interactions $J$ (green circles), $\Gamma$ (blue triangles), and $\Gamma'$ (magenta inverted triangles). It suggests the possible phase transition from a magnetically ordered state to KQSL by controlling the strain.

Second, our results are also directly applicable to topological phase transitions of KQSLs.
This applicability arises from previous results indicating that topological phase transitions of KQSLs under perturbative external fields can be analyzed through symmetry considerations \cite{Noh2024.5}.
Considering perturbative magnetic fields $(\mathbf{h})$ in the $\epsilon$-$KJ\Gamma \Gamma'$ model, the topological invariant of KQSLs ($\nu(\mathbf{h},\epsilon)$) can be characterized by its symmetry properties \cite{Noh2024.5,Hwang2022.1,Imamura2023.5}. 
In Table~\ref{table4}, we list irreducible representations of the physical quantities same as $\nu(\mathbf{h},\epsilon)$, which gives the generic relation,
\small{
\begin{align}
\nu(\mathbf{h},\epsilon)=& \text{Sign}\Big(k_1h_c+k_2h_c^3+k_3h_c(h_a^2+h_b^2)+k_4h_a(h_a^2-3h_b^2)\nonumber \\
&+k_5h_c\epsilon_{a}+k_6h_c\epsilon_{c}+(k_7\mathbf{D}^{\alpha}+k_8\mathbf{D}^{\beta})\times(h_a,h_b)\Big) \label{eqnu} 
\end{align}}
\normalsize{where} $k_{1} \text{--} k_{8}$ are real constants.

Figure~\ref{fig4} illustrates the schematic behavior of the topological invariant under three different strain groups, with the normalized interaction parameters ($k_{1} \text{--} k_{8}=1$). The magnetic field is fixed in the in-plane direction, described by $\mathbf{h} = \cos{\phi} \, \hat{\mathbf{a}} + \sin{\phi} \, \hat{\mathbf{b}}$, where $\phi$ denotes the angle between the field and the $\hat{\mathbf{a}}$-axis.

In the absence of strain, topological phase transitions occur at $\phi = (30 + 60n)^{\circ}$ for $n = 0, 1, 2, 3, 4, 5$, in accordance with previous theoretical and experimental studies~\cite{Noh2024.5,Kitaev2005.10,Hwang2022.1,Imamura2023.5}. These specific angles correspond to high-symmetry directions in the underlying honeycomb lattice. Note that the change of the topological invariant can be induced purely by rotating the magnetic field. 
Now with strains, one can either partially or completely break the crystal symmetry as discussed above, depending on the nature of the deformation. As a result, the topological transition points shift from their pristine positions, reflecting the underlying changes in the spin Hamiltonian. This is clearly observed in Figs.~\ref{fig4}b and \ref {fig4}c, where the location of phase boundaries is altered under different strain configurations.

Most notably, symmetry-protected topological phase transitions—robust even beyond the nearest-neighbor spin model \cite{Winter2018.2}—are indicated by the red points and lines in Fig.~\ref{fig4}.
Figures~\ref{fig4}a and \ref{fig4}b show the crystal structure of $\alpha$-RuCl$_3$ as determined from experimental measurements~\cite{Fletcher1967.1,Kubota2015.3}, consistent with either the $P\bar{3}1m$ (Fig.~\ref{fig4}a) or $C2/m$ (Fig.~\ref{fig4}b) symmetry group. In these configurations, a topological phase transition still occurs when the magnetic field is aligned with the $\hat{\mathbf{b}}$-axis (See red vertical lines).
In contrast, Figure~\ref{fig4}c highlights a striking deviation: the blue line indicates that a topological phase transition can now be induced by strain alone,
 even when the magnetic field is strictly aligned with the $\hat{\mathbf{b}}$-axis. This finding is crucial because it demonstrates the possibility of engineering phase transitions through external mechanical control, independent of field directionality.
We thus propose that the blue line in Fig.~\ref{fig4}c provides a promising experimental pathway for observing strain-driven topological phase transitions not present in Figs.~\ref{fig4}a and \ref{fig4}b. This transition can serve as a diagnostic signature for identifying the KQSL state, offering a new direction for material characterization and manipulation.

\section*{Discussion}
In this work, we develop a fully generalized spin model for candidate materials of KQSLs under homogeneous deformation, labeled $\epsilon$-$KJ\Gamma\Gamma'$.  It generalizes the conventional $KJ\Gamma\Gamma'$ model originally formulated for the high-symmetry limit.
While our study is explicitly grounded in strain effects, the methodology we have developed is not confined to this context alone. 
For example, the breaking of specific rotational or mirror symmetries in the crystal structure under strain was key to generating new anisotropic spin interactions. A similar process can be envisaged under uniaxial pressure that alters hopping integrals and exchange pathways. The symmetry-based decomposition of emergent interactions in our framework ensures that such extensions can be implemented systematically. 
Once the symmetry group of the deformed system is identified, the allowed spin interactions can be classified, and their magnitudes can be evaluated using perturbative or numerical methods. 
Our strategy may also be applicable to a spin model with non nearest-neighbor spin interactions such as the third nearest neighbor interaction ($J_3$) \cite{Winter2018.2} by performing additional higher order calculations. 
Thus, our framework serves not only as a tool for understanding strain effects, but also as a generalizable strategy for mapping out effective spin Hamiltonians in a wide range of materials and experimental conditions.

Furthermore, the implications of our findings are not solely theoretical:
we also estimate the coupling constants that can be measured experimentally.
Under moderate strains (e.g., $3\%$ tensile or compressive strain), the magnitudes of emergent anisotropic interactions become comparable to or even exceed those of the original isotropic terms. 
This indicates that strain effects are not merely a symmetry-tuning-parameter but can serve as a decisive control knob for driving the system into qualitatively new regimes of behavior. For instance, various experiments, including inelastic neutron scattering, can be employed to extract spin-wave spectra or dynamical spin structure factors, which are directly influenced by the values of the spin exchange interactions. Since our model provides explicit numerical estimates of all these couplings under various strains, it facilitates a one-to-one comparison between theoretical predictions and experimental measurements.
We believe that the strain-driven quantum phase transitions of Kitaev materials can be experimentally explored by applying strain effects, for example, epitaxial strain in thin films \cite{Kim2024.7}.
  
Looking forward, several promising directions emerge from our study. First, it would be interesting to investigate time-dependent or dynamic strain effects induced by, e.g., ultrafast optical pulses or acoustic phonons. These could provide a route to dynamically modulate spin interactions and induce transient quantum phases. Second, the extension of our methodology to three-dimensional Kitaev materials or different class of QSLs with different lattice geometries could lead to further discoveries of exotic spin states.


\section*{Method}
To calculate the electronic structures, we performed the first-principles calculations within the DFT framework. For the exchange-correlation energy, we used the generalized gradient approximation (GGA), as parametrized by Perdew, Burke, and Ernzerhof (PBE) \cite{perdew_generalized_1996}. For the pristine $P\bar{3}1m$-RuCl$_3$ monolayer, taking SOC into account, the structure was fully optimized until the residual forces were less than 1\si{\milli\electronvolt/\angstrom} by using the Vienna Ab initio Simulation Package (VASP) code \cite{kresse_efficiency_1996, kresse_efficient_1996}. In this calculation, we adopt 400 \si{\electronvolt} of the plane-wave energy cutoff. For further electronic analyses including the $j_{\text{eff}}$-projected band structure, we used the OpenMX package \cite{ozaki_variationally_2003, kim_electronic_2014}, which adopt the linear combination of pseudo-atomic orbital (LCPAO) basis. Dudarev's rotationally invariant DFT+$U$ formalism with the effective $U_\text{eff}=2$ \si{\electronvolt} was used for on-site Coulomb interactions \cite{dudarev_electron-energy-loss_1998, han_mathrmon_2006}. 500 Ry of the energy cutoff was used for the real-space sampling. We chose $15\times15\times1$ and $13\times7\times1$ $k$-points for the primitive cell and the $1\times\sqrt{3}\times1$ magnetic supercell, respectively. To estimate hopping parameters, we used Wannier function formalism as implemented in Wannier90 code \cite{pizzi_wannier90_2020}. Based on these parameters, we adopted $U = 3$ $\si{\electronvolt}$ and $J_\text{H}/U = 0.15$ \cite{kim_kitaev_2015,Kim2016.4} for calculating spin interactions, which constitute the primary dataset used in our main analysis. From extensive check calculations, we found that the values depend mainly on $J_\text{H}/U$, but the presented results are quite robust in the range of $0.05 < J_\text{H}/U < 0.33$.

\vspace{5mm}




\noindent\textbf{Acknowledgements:} We thank Takasada Shibauchi, Kyusung Hwang, Heung-Sik Kim, and Jae Hoon Kim for their invaluable comments and discussions. This work was supported by 2022M3H4A1A04074153, the National Research Foundation of Korea(NRF) grant funded by the Korea government(MSIT) (Grant Nos. RS-2025-00559042, RS-2023-00253716, and RS-2025-00559286), the Nano \& Material Technology Development Program through  the National Research Foundation of Korea(NRF) funded by Ministry of Science and ICT(RS-2023-00281839, RS-2024-00451261) and National Measurement Standards Services and Technical Support for Industries funded by Korea Research Institute of Standards and Science (KRISS–2025–GP2025-0015).

}
\bibliographystyle{apsrev}
\bibliography{references}

\begin{thebibliography}{76}
\expandafter\ifx\csname natexlab\endcsname\relax\def\natexlab#1{#1}\fi
\expandafter\ifx\csname bibnamefont\endcsname\relax
  \def\bibnamefont#1{#1}\fi
\expandafter\ifx\csname bibfnamefont\endcsname\relax
  \def\bibfnamefont#1{#1}\fi
\expandafter\ifx\csname citenamefont\endcsname\relax
  \def\citenamefont#1{#1}\fi
\expandafter\ifx\csname url\endcsname\relax
  \def\url#1{\texttt{#1}}\fi
\expandafter\ifx\csname urlprefix\endcsname\relax\def\urlprefix{URL }\fi
\providecommand{\bibinfo}[2]{#2}
\providecommand{\eprint}[2][]{\url{#2}}

\bibitem[{\citenamefont{Zhou et~al.}(2017)\citenamefont{Zhou, Kanoda, and
  Ng}}]{Zhou2017.4}
\bibinfo{author}{\bibfnamefont{Y.}~\bibnamefont{Zhou}},
  \bibinfo{author}{\bibfnamefont{K.}~\bibnamefont{Kanoda}}, \bibnamefont{and}
  \bibinfo{author}{\bibfnamefont{T.-K.} \bibnamefont{Ng}},
  \bibinfo{journal}{Rev. Mod. Phys.} \textbf{\bibinfo{volume}{89}},
  \bibinfo{pages}{025003} (\bibinfo{year}{2017}),
  \urlprefix\url{https://link.aps.org/doi/10.1103/RevModPhys.89.025003}.

\bibitem[{\citenamefont{Savary and Balents}(2016)}]{Savary2016.11}
\bibinfo{author}{\bibfnamefont{L.}~\bibnamefont{Savary}} \bibnamefont{and}
  \bibinfo{author}{\bibfnamefont{L.}~\bibnamefont{Balents}},
  \bibinfo{journal}{Reports on Progress in Physics}
  \textbf{\bibinfo{volume}{80}}, \bibinfo{pages}{016502}
  (\bibinfo{year}{2016}),
  \urlprefix\url{https://dx.doi.org/10.1088/0034-4885/80/1/016502}.

\bibitem[{\citenamefont{Balents}(2010)}]{Balents2010.3}
\bibinfo{author}{\bibfnamefont{L.}~\bibnamefont{Balents}},
  \bibinfo{journal}{Nature} \textbf{\bibinfo{volume}{464}},
  \bibinfo{pages}{199} (\bibinfo{year}{2010}), ISSN \bibinfo{issn}{1476-4687},
  \urlprefix\url{https://doi.org/10.1038/nature08917}.

\bibitem[{\citenamefont{Anderson}(1973)}]{Anderson1973.2}
\bibinfo{author}{\bibfnamefont{P.}~\bibnamefont{Anderson}},
  \bibinfo{journal}{Materials Research Bulletin} \textbf{\bibinfo{volume}{8}},
  \bibinfo{pages}{153} (\bibinfo{year}{1973}), ISSN \bibinfo{issn}{0025-5408},
  \urlprefix\url{https://www.sciencedirect.com/science/article/pii/0025540873901670}.

\bibitem[{\citenamefont{Knolle and Moessner}(2019)}]{Knolle2019.3}
\bibinfo{author}{\bibfnamefont{J.}~\bibnamefont{Knolle}} \bibnamefont{and}
  \bibinfo{author}{\bibfnamefont{R.}~\bibnamefont{Moessner}},
  \bibinfo{journal}{Annual Review of Condensed Matter Physics}
  \textbf{\bibinfo{volume}{10}}, \bibinfo{pages}{451} (\bibinfo{year}{2019}),
  \eprint{https://doi.org/10.1146/annurev-conmatphys-031218-013401},
  \urlprefix\url{https://doi.org/10.1146/annurev-conmatphys-031218-013401}.

\bibitem[{\citenamefont{Nayak et~al.}(2008)\citenamefont{Nayak, Simon, Stern,
  Freedman, and Das~Sarma}}]{Nayak2008.9}
\bibinfo{author}{\bibfnamefont{C.}~\bibnamefont{Nayak}},
  \bibinfo{author}{\bibfnamefont{S.~H.} \bibnamefont{Simon}},
  \bibinfo{author}{\bibfnamefont{A.}~\bibnamefont{Stern}},
  \bibinfo{author}{\bibfnamefont{M.}~\bibnamefont{Freedman}}, \bibnamefont{and}
  \bibinfo{author}{\bibfnamefont{S.}~\bibnamefont{Das~Sarma}},
  \bibinfo{journal}{Rev. Mod. Phys.} \textbf{\bibinfo{volume}{80}},
  \bibinfo{pages}{1083} (\bibinfo{year}{2008}),
  \urlprefix\url{https://link.aps.org/doi/10.1103/RevModPhys.80.1083}.

\bibitem[{\citenamefont{Kitaev}(2003)}]{Kitaev2002.5}
\bibinfo{author}{\bibfnamefont{A.}~\bibnamefont{Kitaev}},
  \bibinfo{journal}{Annals of Physics} \textbf{\bibinfo{volume}{303}},
  \bibinfo{pages}{2} (\bibinfo{year}{2003}), ISSN \bibinfo{issn}{0003-4916},
  \urlprefix\url{https://www.sciencedirect.com/science/article/pii/S0003491602000180}.

\bibitem[{\citenamefont{Kitaev}(2006)}]{Kitaev2005.10}
\bibinfo{author}{\bibfnamefont{A.}~\bibnamefont{Kitaev}},
  \bibinfo{journal}{Annals of Physics} \textbf{\bibinfo{volume}{321}},
  \bibinfo{pages}{2} (\bibinfo{year}{2006}), ISSN \bibinfo{issn}{0003-4916},
  \bibinfo{note}{january Special Issue},
  \urlprefix\url{https://www.sciencedirect.com/science/article/pii/S0003491605002381}.

\bibitem[{\citenamefont{Jackeli and Khaliullin}(2009)}]{Jackeli2009.1}
\bibinfo{author}{\bibfnamefont{G.}~\bibnamefont{Jackeli}} \bibnamefont{and}
  \bibinfo{author}{\bibfnamefont{G.}~\bibnamefont{Khaliullin}},
  \bibinfo{journal}{Phys. Rev. Lett.} \textbf{\bibinfo{volume}{102}},
  \bibinfo{pages}{017205} (\bibinfo{year}{2009}),
  \urlprefix\url{https://link.aps.org/doi/10.1103/PhysRevLett.102.017205}.

\bibitem[{\citenamefont{Chaloupka et~al.}(2010)\citenamefont{Chaloupka,
  Jackeli, and Khaliullin}}]{Chaloupka2010.7}
\bibinfo{author}{\bibfnamefont{J.~c.~v.} \bibnamefont{Chaloupka}},
  \bibinfo{author}{\bibfnamefont{G.}~\bibnamefont{Jackeli}}, \bibnamefont{and}
  \bibinfo{author}{\bibfnamefont{G.}~\bibnamefont{Khaliullin}},
  \bibinfo{journal}{Phys. Rev. Lett.} \textbf{\bibinfo{volume}{105}},
  \bibinfo{pages}{027204} (\bibinfo{year}{2010}),
  \urlprefix\url{https://link.aps.org/doi/10.1103/PhysRevLett.105.027204}.

\bibitem[{\citenamefont{Rau et~al.}(2014)\citenamefont{Rau, Lee, and
  Kee}}]{Rau2014.2}
\bibinfo{author}{\bibfnamefont{J.~G.} \bibnamefont{Rau}},
  \bibinfo{author}{\bibfnamefont{E.~K.-H.} \bibnamefont{Lee}},
  \bibnamefont{and} \bibinfo{author}{\bibfnamefont{H.-Y.} \bibnamefont{Kee}},
  \bibinfo{journal}{Phys. Rev. Lett.} \textbf{\bibinfo{volume}{112}},
  \bibinfo{pages}{077204} (\bibinfo{year}{2014}),
  \urlprefix\url{https://link.aps.org/doi/10.1103/PhysRevLett.112.077204}.

\bibitem[{\citenamefont{Rau and Kee}(2014)}]{rau_trigonal_2014}
\bibinfo{author}{\bibfnamefont{J.~G.} \bibnamefont{Rau}} \bibnamefont{and}
  \bibinfo{author}{\bibfnamefont{H.-Y.} \bibnamefont{Kee}},
  \emph{\bibinfo{title}{Trigonal distortion in the honeycomb iridates:
  {Proximity} of zigzag and spiral phases in
  $\mathrm{Na}_{2}$$\mathrm{IrO}_{3}$}} (\bibinfo{year}{2014}),
  \bibinfo{note}{arXiv:1408.4811 [cond-mat]},
  \urlprefix\url{http://arxiv.org/abs/1408.4811}.

\bibitem[{\citenamefont{Plumb et~al.}(2014)\citenamefont{Plumb, Clancy,
  Sandilands, Shankar, Hu, Burch, Kee, and Kim}}]{Plumb2014.7}
\bibinfo{author}{\bibfnamefont{K.~W.} \bibnamefont{Plumb}},
  \bibinfo{author}{\bibfnamefont{J.~P.} \bibnamefont{Clancy}},
  \bibinfo{author}{\bibfnamefont{L.~J.} \bibnamefont{Sandilands}},
  \bibinfo{author}{\bibfnamefont{V.~V.} \bibnamefont{Shankar}},
  \bibinfo{author}{\bibfnamefont{Y.~F.} \bibnamefont{Hu}},
  \bibinfo{author}{\bibfnamefont{K.~S.} \bibnamefont{Burch}},
  \bibinfo{author}{\bibfnamefont{H.-Y.} \bibnamefont{Kee}}, \bibnamefont{and}
  \bibinfo{author}{\bibfnamefont{Y.-J.} \bibnamefont{Kim}},
  \bibinfo{journal}{Phys. Rev. B} \textbf{\bibinfo{volume}{90}},
  \bibinfo{pages}{041112} (\bibinfo{year}{2014}),
  \urlprefix\url{https://link.aps.org/doi/10.1103/PhysRevB.90.041112}.

\bibitem[{\citenamefont{Koitzsch et~al.}(2016)\citenamefont{Koitzsch,
  Habenicht, M\"uller, Knupfer, B\"uchner, Kandpal, van~den Brink, Nowak,
  Isaeva, and Doert}}]{Koitzsch2016.9}
\bibinfo{author}{\bibfnamefont{A.}~\bibnamefont{Koitzsch}},
  \bibinfo{author}{\bibfnamefont{C.}~\bibnamefont{Habenicht}},
  \bibinfo{author}{\bibfnamefont{E.}~\bibnamefont{M\"uller}},
  \bibinfo{author}{\bibfnamefont{M.}~\bibnamefont{Knupfer}},
  \bibinfo{author}{\bibfnamefont{B.}~\bibnamefont{B\"uchner}},
  \bibinfo{author}{\bibfnamefont{H.~C.} \bibnamefont{Kandpal}},
  \bibinfo{author}{\bibfnamefont{J.}~\bibnamefont{van~den Brink}},
  \bibinfo{author}{\bibfnamefont{D.}~\bibnamefont{Nowak}},
  \bibinfo{author}{\bibfnamefont{A.}~\bibnamefont{Isaeva}}, \bibnamefont{and}
  \bibinfo{author}{\bibfnamefont{T.}~\bibnamefont{Doert}},
  \bibinfo{journal}{Phys. Rev. Lett.} \textbf{\bibinfo{volume}{117}},
  \bibinfo{pages}{126403} (\bibinfo{year}{2016}),
  \urlprefix\url{https://link.aps.org/doi/10.1103/PhysRevLett.117.126403}.

\bibitem[{\citenamefont{Sandilands et~al.}(2015)\citenamefont{Sandilands, Tian,
  Plumb, Kim, and Burch}}]{Sandilands2015.4}
\bibinfo{author}{\bibfnamefont{L.~J.} \bibnamefont{Sandilands}},
  \bibinfo{author}{\bibfnamefont{Y.}~\bibnamefont{Tian}},
  \bibinfo{author}{\bibfnamefont{K.~W.} \bibnamefont{Plumb}},
  \bibinfo{author}{\bibfnamefont{Y.-J.} \bibnamefont{Kim}}, \bibnamefont{and}
  \bibinfo{author}{\bibfnamefont{K.~S.} \bibnamefont{Burch}},
  \bibinfo{journal}{Phys. Rev. Lett.} \textbf{\bibinfo{volume}{114}},
  \bibinfo{pages}{147201} (\bibinfo{year}{2015}),
  \urlprefix\url{https://link.aps.org/doi/10.1103/PhysRevLett.114.147201}.

\bibitem[{\citenamefont{Jiang et~al.}(2011)\citenamefont{Jiang, Gu, Qi, and
  Trebst}}]{Kim2015.6}
\bibinfo{author}{\bibfnamefont{H.-C.} \bibnamefont{Jiang}},
  \bibinfo{author}{\bibfnamefont{Z.-C.} \bibnamefont{Gu}},
  \bibinfo{author}{\bibfnamefont{X.-L.} \bibnamefont{Qi}}, \bibnamefont{and}
  \bibinfo{author}{\bibfnamefont{S.}~\bibnamefont{Trebst}},
  \bibinfo{journal}{Phys. Rev. B} \textbf{\bibinfo{volume}{83}},
  \bibinfo{pages}{245104} (\bibinfo{year}{2011}),
  \urlprefix\url{https://link.aps.org/doi/10.1103/PhysRevB.83.245104}.

\bibitem[{\citenamefont{Kim and Kee}(2016)}]{Kim2016.4}
\bibinfo{author}{\bibfnamefont{H.-S.} \bibnamefont{Kim}} \bibnamefont{and}
  \bibinfo{author}{\bibfnamefont{H.-Y.} \bibnamefont{Kee}},
  \bibinfo{journal}{Phys. Rev. B} \textbf{\bibinfo{volume}{93}},
  \bibinfo{pages}{155143} (\bibinfo{year}{2016}),
  \urlprefix\url{https://link.aps.org/doi/10.1103/PhysRevB.93.155143}.

\bibitem[{\citenamefont{Winter et~al.}(2017)\citenamefont{Winter, Tsirlin,
  Daghofer, van~den Brink, Singh, Gegenwart, and Valentí}}]{Winter2017.11}
\bibinfo{author}{\bibfnamefont{S.~M.} \bibnamefont{Winter}},
  \bibinfo{author}{\bibfnamefont{A.~A.} \bibnamefont{Tsirlin}},
  \bibinfo{author}{\bibfnamefont{M.}~\bibnamefont{Daghofer}},
  \bibinfo{author}{\bibfnamefont{J.}~\bibnamefont{van~den Brink}},
  \bibinfo{author}{\bibfnamefont{Y.}~\bibnamefont{Singh}},
  \bibinfo{author}{\bibfnamefont{P.}~\bibnamefont{Gegenwart}},
  \bibnamefont{and} \bibinfo{author}{\bibfnamefont{R.}~\bibnamefont{Valentí}},
  \bibinfo{journal}{Journal of Physics: Condensed Matter}
  \textbf{\bibinfo{volume}{29}}, \bibinfo{pages}{493002}
  (\bibinfo{year}{2017}),
  \urlprefix\url{https://dx.doi.org/10.1088/1361-648X/aa8cf5}.

\bibitem[{\citenamefont{Winter et~al.}(2016)\citenamefont{Winter, Li, Jeschke,
  and Valent\'{\i}}}]{Winter2016.6}
\bibinfo{author}{\bibfnamefont{S.~M.} \bibnamefont{Winter}},
  \bibinfo{author}{\bibfnamefont{Y.}~\bibnamefont{Li}},
  \bibinfo{author}{\bibfnamefont{H.~O.} \bibnamefont{Jeschke}},
  \bibnamefont{and}
  \bibinfo{author}{\bibfnamefont{R.}~\bibnamefont{Valent\'{\i}}},
  \bibinfo{journal}{Phys. Rev. B} \textbf{\bibinfo{volume}{93}},
  \bibinfo{pages}{214431} (\bibinfo{year}{2016}),
  \urlprefix\url{https://link.aps.org/doi/10.1103/PhysRevB.93.214431}.

\bibitem[{\citenamefont{Winter et~al.}(2018)\citenamefont{Winter, Riedl, Kaib,
  Coldea, and Valent\'{\i}}}]{Winter2018.2}
\bibinfo{author}{\bibfnamefont{S.~M.} \bibnamefont{Winter}},
  \bibinfo{author}{\bibfnamefont{K.}~\bibnamefont{Riedl}},
  \bibinfo{author}{\bibfnamefont{D.}~\bibnamefont{Kaib}},
  \bibinfo{author}{\bibfnamefont{R.}~\bibnamefont{Coldea}}, \bibnamefont{and}
  \bibinfo{author}{\bibfnamefont{R.}~\bibnamefont{Valent\'{\i}}},
  \bibinfo{journal}{Phys. Rev. Lett.} \textbf{\bibinfo{volume}{120}},
  \bibinfo{pages}{077203} (\bibinfo{year}{2018}),
  \urlprefix\url{https://link.aps.org/doi/10.1103/PhysRevLett.120.077203}.

\bibitem[{\citenamefont{Yadav et~al.}(2016)\citenamefont{Yadav, Bogdanov,
  Katukuri, Nishimoto, van~den Brink, and Hozoi}}]{Yadav2016.11}
\bibinfo{author}{\bibfnamefont{R.}~\bibnamefont{Yadav}},
  \bibinfo{author}{\bibfnamefont{N.~A.} \bibnamefont{Bogdanov}},
  \bibinfo{author}{\bibfnamefont{V.~M.} \bibnamefont{Katukuri}},
  \bibinfo{author}{\bibfnamefont{S.}~\bibnamefont{Nishimoto}},
  \bibinfo{author}{\bibfnamefont{J.}~\bibnamefont{van~den Brink}},
  \bibnamefont{and} \bibinfo{author}{\bibfnamefont{L.}~\bibnamefont{Hozoi}},
  \bibinfo{journal}{Scientific Reports} \textbf{\bibinfo{volume}{6}},
  \bibinfo{pages}{37925} (\bibinfo{year}{2016}), ISSN
  \bibinfo{issn}{2045-2322}, \urlprefix\url{https://doi.org/10.1038/srep37925}.

\bibitem[{\citenamefont{Takeda et~al.}(2022)\citenamefont{Takeda, Mai, Akazawa,
  Tamura, Yan, Moovendaran, Raju, Sankar, Choi, and Yamashita}}]{Takeda2022.11}
\bibinfo{author}{\bibfnamefont{H.}~\bibnamefont{Takeda}},
  \bibinfo{author}{\bibfnamefont{J.}~\bibnamefont{Mai}},
  \bibinfo{author}{\bibfnamefont{M.}~\bibnamefont{Akazawa}},
  \bibinfo{author}{\bibfnamefont{K.}~\bibnamefont{Tamura}},
  \bibinfo{author}{\bibfnamefont{J.}~\bibnamefont{Yan}},
  \bibinfo{author}{\bibfnamefont{K.}~\bibnamefont{Moovendaran}},
  \bibinfo{author}{\bibfnamefont{K.}~\bibnamefont{Raju}},
  \bibinfo{author}{\bibfnamefont{R.}~\bibnamefont{Sankar}},
  \bibinfo{author}{\bibfnamefont{K.-Y.} \bibnamefont{Choi}}, \bibnamefont{and}
  \bibinfo{author}{\bibfnamefont{M.}~\bibnamefont{Yamashita}},
  \bibinfo{journal}{Phys. Rev. Res.} \textbf{\bibinfo{volume}{4}},
  \bibinfo{pages}{L042035} (\bibinfo{year}{2022}),
  \urlprefix\url{https://link.aps.org/doi/10.1103/PhysRevResearch.4.L042035}.

\bibitem[{\citenamefont{Viciu et~al.}(2007)\citenamefont{Viciu, Huang, Morosan,
  Zandbergen, Greenbaum, McQueen, and Cava}}]{Viciu2007.1}
\bibinfo{author}{\bibfnamefont{L.}~\bibnamefont{Viciu}},
  \bibinfo{author}{\bibfnamefont{Q.}~\bibnamefont{Huang}},
  \bibinfo{author}{\bibfnamefont{E.}~\bibnamefont{Morosan}},
  \bibinfo{author}{\bibfnamefont{H.}~\bibnamefont{Zandbergen}},
  \bibinfo{author}{\bibfnamefont{N.}~\bibnamefont{Greenbaum}},
  \bibinfo{author}{\bibfnamefont{T.}~\bibnamefont{McQueen}}, \bibnamefont{and}
  \bibinfo{author}{\bibfnamefont{R.}~\bibnamefont{Cava}},
  \bibinfo{journal}{Journal of Solid State Chemistry}
  \textbf{\bibinfo{volume}{180}}, \bibinfo{pages}{1060} (\bibinfo{year}{2007}),
  ISSN \bibinfo{issn}{0022-4596},
  \urlprefix\url{https://www.sciencedirect.com/science/article/pii/S0022459607000175}.

\bibitem[{\citenamefont{Songvilay et~al.}(2020)\citenamefont{Songvilay, Robert,
  Petit, Rodriguez-Rivera, Ratcliff, Damay, Bal\'edent, Jim\'enez-Ruiz, Lejay,
  Pachoud et~al.}}]{Songvilay2020.12}
\bibinfo{author}{\bibfnamefont{M.}~\bibnamefont{Songvilay}},
  \bibinfo{author}{\bibfnamefont{J.}~\bibnamefont{Robert}},
  \bibinfo{author}{\bibfnamefont{S.}~\bibnamefont{Petit}},
  \bibinfo{author}{\bibfnamefont{J.~A.} \bibnamefont{Rodriguez-Rivera}},
  \bibinfo{author}{\bibfnamefont{W.~D.} \bibnamefont{Ratcliff}},
  \bibinfo{author}{\bibfnamefont{F.}~\bibnamefont{Damay}},
  \bibinfo{author}{\bibfnamefont{V.}~\bibnamefont{Bal\'edent}},
  \bibinfo{author}{\bibfnamefont{M.}~\bibnamefont{Jim\'enez-Ruiz}},
  \bibinfo{author}{\bibfnamefont{P.}~\bibnamefont{Lejay}},
  \bibinfo{author}{\bibfnamefont{E.}~\bibnamefont{Pachoud}},
  \bibnamefont{et~al.}, \bibinfo{journal}{Phys. Rev. B}
  \textbf{\bibinfo{volume}{102}}, \bibinfo{pages}{224429}
  (\bibinfo{year}{2020}),
  \urlprefix\url{https://link.aps.org/doi/10.1103/PhysRevB.102.224429}.

\bibitem[{\citenamefont{Lin et~al.}(2021)\citenamefont{Lin, Jeong, Kim, Wang,
  Huang, Masuda, Asai, Itoh, Günther, Russina et~al.}}]{Lin2021.9}
\bibinfo{author}{\bibfnamefont{G.}~\bibnamefont{Lin}},
  \bibinfo{author}{\bibfnamefont{J.}~\bibnamefont{Jeong}},
  \bibinfo{author}{\bibfnamefont{C.}~\bibnamefont{Kim}},
  \bibinfo{author}{\bibfnamefont{Y.}~\bibnamefont{Wang}},
  \bibinfo{author}{\bibfnamefont{Q.}~\bibnamefont{Huang}},
  \bibinfo{author}{\bibfnamefont{T.}~\bibnamefont{Masuda}},
  \bibinfo{author}{\bibfnamefont{S.}~\bibnamefont{Asai}},
  \bibinfo{author}{\bibfnamefont{S.}~\bibnamefont{Itoh}},
  \bibinfo{author}{\bibfnamefont{G.}~\bibnamefont{Günther}},
  \bibinfo{author}{\bibfnamefont{M.}~\bibnamefont{Russina}},
  \bibnamefont{et~al.}, \bibinfo{journal}{Nature Communications}
  \textbf{\bibinfo{volume}{12}}, \bibinfo{pages}{5559} (\bibinfo{year}{2021}),
  ISSN \bibinfo{issn}{2041-1723},
  \urlprefix\url{https://doi.org/10.1038/s41467-021-25567-7}.

\bibitem[{\citenamefont{Wulferding et~al.}(2020)\citenamefont{Wulferding, Choi,
  Do, Lee, Lemmens, Faugeras, Gallais, and Choi}}]{Wulferding2020.3}
\bibinfo{author}{\bibfnamefont{D.}~\bibnamefont{Wulferding}},
  \bibinfo{author}{\bibfnamefont{Y.}~\bibnamefont{Choi}},
  \bibinfo{author}{\bibfnamefont{S.-H.} \bibnamefont{Do}},
  \bibinfo{author}{\bibfnamefont{C.~H.} \bibnamefont{Lee}},
  \bibinfo{author}{\bibfnamefont{P.}~\bibnamefont{Lemmens}},
  \bibinfo{author}{\bibfnamefont{C.}~\bibnamefont{Faugeras}},
  \bibinfo{author}{\bibfnamefont{Y.}~\bibnamefont{Gallais}}, \bibnamefont{and}
  \bibinfo{author}{\bibfnamefont{K.-Y.} \bibnamefont{Choi}},
  \bibinfo{journal}{Nature Communications} \textbf{\bibinfo{volume}{11}},
  \bibinfo{pages}{1603} (\bibinfo{year}{2020}), ISSN \bibinfo{issn}{2041-1723},
  \urlprefix\url{https://doi.org/10.1038/s41467-020-15370-1}.

\bibitem[{\citenamefont{Tanaka et~al.}(2022)\citenamefont{Tanaka, Mizukami,
  Harasawa, Hashimoto, Hwang, Kurita, Tanaka, Fujimoto, Matsuda, Moon
  et~al.}}]{Tanaka2022.1}
\bibinfo{author}{\bibfnamefont{O.}~\bibnamefont{Tanaka}},
  \bibinfo{author}{\bibfnamefont{Y.}~\bibnamefont{Mizukami}},
  \bibinfo{author}{\bibfnamefont{R.}~\bibnamefont{Harasawa}},
  \bibinfo{author}{\bibfnamefont{K.}~\bibnamefont{Hashimoto}},
  \bibinfo{author}{\bibfnamefont{K.}~\bibnamefont{Hwang}},
  \bibinfo{author}{\bibfnamefont{N.}~\bibnamefont{Kurita}},
  \bibinfo{author}{\bibfnamefont{H.}~\bibnamefont{Tanaka}},
  \bibinfo{author}{\bibfnamefont{S.}~\bibnamefont{Fujimoto}},
  \bibinfo{author}{\bibfnamefont{Y.}~\bibnamefont{Matsuda}},
  \bibinfo{author}{\bibfnamefont{E.~G.} \bibnamefont{Moon}},
  \bibnamefont{et~al.}, \bibinfo{journal}{Nature Physics}
  \textbf{\bibinfo{volume}{18}}, \bibinfo{pages}{429} (\bibinfo{year}{2022}),
  ISSN \bibinfo{issn}{1745-2481},
  \urlprefix\url{https://doi.org/10.1038/s41567-021-01488-6}.

\bibitem[{\citenamefont{Xing et~al.}(2025)\citenamefont{Xing, Namba, Imamura,
  Ishihara, Suetsugu, Asaba, Hashimoto, Shibauchi, Matsuda, and
  Kasahara}}]{Xing2025.3}
\bibinfo{author}{\bibfnamefont{Y.}~\bibnamefont{Xing}},
  \bibinfo{author}{\bibfnamefont{R.}~\bibnamefont{Namba}},
  \bibinfo{author}{\bibfnamefont{K.}~\bibnamefont{Imamura}},
  \bibinfo{author}{\bibfnamefont{K.}~\bibnamefont{Ishihara}},
  \bibinfo{author}{\bibfnamefont{S.}~\bibnamefont{Suetsugu}},
  \bibinfo{author}{\bibfnamefont{T.}~\bibnamefont{Asaba}},
  \bibinfo{author}{\bibfnamefont{K.}~\bibnamefont{Hashimoto}},
  \bibinfo{author}{\bibfnamefont{T.}~\bibnamefont{Shibauchi}},
  \bibinfo{author}{\bibfnamefont{Y.}~\bibnamefont{Matsuda}}, \bibnamefont{and}
  \bibinfo{author}{\bibfnamefont{Y.}~\bibnamefont{Kasahara}},
  \bibinfo{journal}{npj Quantum Materials} \textbf{\bibinfo{volume}{10}},
  \bibinfo{pages}{33} (\bibinfo{year}{2025}), ISSN \bibinfo{issn}{2397-4648},
  \urlprefix\url{https://doi.org/10.1038/s41535-025-00749-4}.

\bibitem[{\citenamefont{Kasahara et~al.}(2022)\citenamefont{Kasahara, Suetsugu,
  Asaba, Kasahara, Shibauchi, Kurita, Tanaka, and Matsuda}}]{Kasahara2022.8}
\bibinfo{author}{\bibfnamefont{Y.}~\bibnamefont{Kasahara}},
  \bibinfo{author}{\bibfnamefont{S.}~\bibnamefont{Suetsugu}},
  \bibinfo{author}{\bibfnamefont{T.}~\bibnamefont{Asaba}},
  \bibinfo{author}{\bibfnamefont{S.}~\bibnamefont{Kasahara}},
  \bibinfo{author}{\bibfnamefont{T.}~\bibnamefont{Shibauchi}},
  \bibinfo{author}{\bibfnamefont{N.}~\bibnamefont{Kurita}},
  \bibinfo{author}{\bibfnamefont{H.}~\bibnamefont{Tanaka}}, \bibnamefont{and}
  \bibinfo{author}{\bibfnamefont{Y.}~\bibnamefont{Matsuda}},
  \bibinfo{journal}{Phys. Rev. B} \textbf{\bibinfo{volume}{106}},
  \bibinfo{pages}{L060410} (\bibinfo{year}{2022}),
  \urlprefix\url{https://link.aps.org/doi/10.1103/PhysRevB.106.L060410}.

\bibitem[{\citenamefont{Kasahara
  et~al.}(2018{\natexlab{a}})\citenamefont{Kasahara, Ohnishi, Mizukami, Tanaka,
  Ma, Sugii, Kurita, Tanaka, Nasu, Motome et~al.}}]{Kasahara2018.7}
\bibinfo{author}{\bibfnamefont{Y.}~\bibnamefont{Kasahara}},
  \bibinfo{author}{\bibfnamefont{T.}~\bibnamefont{Ohnishi}},
  \bibinfo{author}{\bibfnamefont{Y.}~\bibnamefont{Mizukami}},
  \bibinfo{author}{\bibfnamefont{O.}~\bibnamefont{Tanaka}},
  \bibinfo{author}{\bibfnamefont{S.}~\bibnamefont{Ma}},
  \bibinfo{author}{\bibfnamefont{K.}~\bibnamefont{Sugii}},
  \bibinfo{author}{\bibfnamefont{N.}~\bibnamefont{Kurita}},
  \bibinfo{author}{\bibfnamefont{H.}~\bibnamefont{Tanaka}},
  \bibinfo{author}{\bibfnamefont{J.}~\bibnamefont{Nasu}},
  \bibinfo{author}{\bibfnamefont{Y.}~\bibnamefont{Motome}},
  \bibnamefont{et~al.}, \bibinfo{journal}{Nature}
  \textbf{\bibinfo{volume}{559}}, \bibinfo{pages}{227}
  (\bibinfo{year}{2018}{\natexlab{a}}), ISSN \bibinfo{issn}{1476-4687},
  \urlprefix\url{https://doi.org/10.1038/s41586-018-0274-0}.

\bibitem[{\citenamefont{Yokoi et~al.}(2021)\citenamefont{Yokoi, Ma, Kasahara,
  Kasahara, Shibauchi, Kurita, Tanaka, Nasu, Motome, Hickey
  et~al.}}]{Yokoi2021.7}
\bibinfo{author}{\bibfnamefont{T.}~\bibnamefont{Yokoi}},
  \bibinfo{author}{\bibfnamefont{S.}~\bibnamefont{Ma}},
  \bibinfo{author}{\bibfnamefont{Y.}~\bibnamefont{Kasahara}},
  \bibinfo{author}{\bibfnamefont{S.}~\bibnamefont{Kasahara}},
  \bibinfo{author}{\bibfnamefont{T.}~\bibnamefont{Shibauchi}},
  \bibinfo{author}{\bibfnamefont{N.}~\bibnamefont{Kurita}},
  \bibinfo{author}{\bibfnamefont{H.}~\bibnamefont{Tanaka}},
  \bibinfo{author}{\bibfnamefont{J.}~\bibnamefont{Nasu}},
  \bibinfo{author}{\bibfnamefont{Y.}~\bibnamefont{Motome}},
  \bibinfo{author}{\bibfnamefont{C.}~\bibnamefont{Hickey}},
  \bibnamefont{et~al.}, \bibinfo{journal}{Science}
  \textbf{\bibinfo{volume}{373}}, \bibinfo{pages}{568} (\bibinfo{year}{2021}),
  \eprint{https://www.science.org/doi/pdf/10.1126/science.aay5551},
  \urlprefix\url{https://www.science.org/doi/abs/10.1126/science.aay5551}.

\bibitem[{\citenamefont{Czajka et~al.}(2021)\citenamefont{Czajka, Gao,
  Hirschberger, Lampen-Kelley, Banerjee, Yan, Mandrus, Nagler, and
  Ong}}]{Czajka2021.5}
\bibinfo{author}{\bibfnamefont{P.}~\bibnamefont{Czajka}},
  \bibinfo{author}{\bibfnamefont{T.}~\bibnamefont{Gao}},
  \bibinfo{author}{\bibfnamefont{M.}~\bibnamefont{Hirschberger}},
  \bibinfo{author}{\bibfnamefont{P.}~\bibnamefont{Lampen-Kelley}},
  \bibinfo{author}{\bibfnamefont{A.}~\bibnamefont{Banerjee}},
  \bibinfo{author}{\bibfnamefont{J.}~\bibnamefont{Yan}},
  \bibinfo{author}{\bibfnamefont{D.~G.} \bibnamefont{Mandrus}},
  \bibinfo{author}{\bibfnamefont{S.~E.} \bibnamefont{Nagler}},
  \bibnamefont{and} \bibinfo{author}{\bibfnamefont{N.~P.} \bibnamefont{Ong}},
  \bibinfo{journal}{Nature Physics} \textbf{\bibinfo{volume}{17}},
  \bibinfo{pages}{915} (\bibinfo{year}{2021}), ISSN \bibinfo{issn}{1745-2481},
  \urlprefix\url{https://doi.org/10.1038/s41567-021-01243-x}.

\bibitem[{\citenamefont{Takagi et~al.}(2019)\citenamefont{Takagi, Takayama,
  Jackeli, Khaliullin, and Nagler}}]{Takagi2019.3}
\bibinfo{author}{\bibfnamefont{H.}~\bibnamefont{Takagi}},
  \bibinfo{author}{\bibfnamefont{T.}~\bibnamefont{Takayama}},
  \bibinfo{author}{\bibfnamefont{G.}~\bibnamefont{Jackeli}},
  \bibinfo{author}{\bibfnamefont{G.}~\bibnamefont{Khaliullin}},
  \bibnamefont{and} \bibinfo{author}{\bibfnamefont{S.~E.}
  \bibnamefont{Nagler}}, \bibinfo{journal}{Nature Reviews Physics}
  \textbf{\bibinfo{volume}{1}}, \bibinfo{pages}{264} (\bibinfo{year}{2019}),
  ISSN \bibinfo{issn}{2522-5820},
  \urlprefix\url{https://doi.org/10.1038/s42254-019-0038-2}.

\bibitem[{\citenamefont{Lefran\ifmmode~\mbox{\c{c}}\else \c{c}\fi{}ois
  et~al.}(2022)\citenamefont{Lefran\ifmmode~\mbox{\c{c}}\else \c{c}\fi{}ois,
  Grissonnanche, Baglo, Lampen-Kelley, Yan, Balz, Mandrus, Nagler, Kim, Kim
  et~al.}}]{Lefrancois2022.4}
\bibinfo{author}{\bibfnamefont{E.}~\bibnamefont{Lefran\ifmmode~\mbox{\c{c}}\else
  \c{c}\fi{}ois}},
  \bibinfo{author}{\bibfnamefont{G.}~\bibnamefont{Grissonnanche}},
  \bibinfo{author}{\bibfnamefont{J.}~\bibnamefont{Baglo}},
  \bibinfo{author}{\bibfnamefont{P.}~\bibnamefont{Lampen-Kelley}},
  \bibinfo{author}{\bibfnamefont{J.-Q.} \bibnamefont{Yan}},
  \bibinfo{author}{\bibfnamefont{C.}~\bibnamefont{Balz}},
  \bibinfo{author}{\bibfnamefont{D.}~\bibnamefont{Mandrus}},
  \bibinfo{author}{\bibfnamefont{S.~E.} \bibnamefont{Nagler}},
  \bibinfo{author}{\bibfnamefont{S.}~\bibnamefont{Kim}},
  \bibinfo{author}{\bibfnamefont{Y.-J.} \bibnamefont{Kim}},
  \bibnamefont{et~al.}, \bibinfo{journal}{Phys. Rev. X}
  \textbf{\bibinfo{volume}{12}}, \bibinfo{pages}{021025}
  (\bibinfo{year}{2022}),
  \urlprefix\url{https://link.aps.org/doi/10.1103/PhysRevX.12.021025}.

\bibitem[{\citenamefont{Rousochatzakis
  et~al.}(2015)\citenamefont{Rousochatzakis, Reuther, Thomale, Rachel, and
  Perkins}}]{Rousochatzakis2015.12}
\bibinfo{author}{\bibfnamefont{I.}~\bibnamefont{Rousochatzakis}},
  \bibinfo{author}{\bibfnamefont{J.}~\bibnamefont{Reuther}},
  \bibinfo{author}{\bibfnamefont{R.}~\bibnamefont{Thomale}},
  \bibinfo{author}{\bibfnamefont{S.}~\bibnamefont{Rachel}}, \bibnamefont{and}
  \bibinfo{author}{\bibfnamefont{N.~B.} \bibnamefont{Perkins}},
  \bibinfo{journal}{Phys. Rev. X} \textbf{\bibinfo{volume}{5}},
  \bibinfo{pages}{041035} (\bibinfo{year}{2015}),
  \urlprefix\url{https://link.aps.org/doi/10.1103/PhysRevX.5.041035}.

\bibitem[{\citenamefont{Hal\'asz et~al.}(2016)\citenamefont{Hal\'asz, Perkins,
  and van~den Brink}}]{Halasz2016.9}
\bibinfo{author}{\bibfnamefont{G.~B.} \bibnamefont{Hal\'asz}},
  \bibinfo{author}{\bibfnamefont{N.~B.} \bibnamefont{Perkins}},
  \bibnamefont{and} \bibinfo{author}{\bibfnamefont{J.}~\bibnamefont{van~den
  Brink}}, \bibinfo{journal}{Phys. Rev. Lett.} \textbf{\bibinfo{volume}{117}},
  \bibinfo{pages}{127203} (\bibinfo{year}{2016}),
  \urlprefix\url{https://link.aps.org/doi/10.1103/PhysRevLett.117.127203}.

\bibitem[{\citenamefont{Ronquillo et~al.}(2019)\citenamefont{Ronquillo, Vengal,
  and Trivedi}}]{Ronquillo2019.4}
\bibinfo{author}{\bibfnamefont{D.~C.} \bibnamefont{Ronquillo}},
  \bibinfo{author}{\bibfnamefont{A.}~\bibnamefont{Vengal}}, \bibnamefont{and}
  \bibinfo{author}{\bibfnamefont{N.}~\bibnamefont{Trivedi}},
  \bibinfo{journal}{Phys. Rev. B} \textbf{\bibinfo{volume}{99}},
  \bibinfo{pages}{140413} (\bibinfo{year}{2019}),
  \urlprefix\url{https://link.aps.org/doi/10.1103/PhysRevB.99.140413}.

\bibitem[{\citenamefont{Patel and Trivedi}(2019)}]{Patel2019.5}
\bibinfo{author}{\bibfnamefont{N.~D.} \bibnamefont{Patel}} \bibnamefont{and}
  \bibinfo{author}{\bibfnamefont{N.}~\bibnamefont{Trivedi}},
  \bibinfo{journal}{Proceedings of the National Academy of Sciences}
  \textbf{\bibinfo{volume}{116}}, \bibinfo{pages}{12199}
  (\bibinfo{year}{2019}),
  \eprint{https://www.pnas.org/doi/pdf/10.1073/pnas.1821406116},
  \urlprefix\url{https://www.pnas.org/doi/abs/10.1073/pnas.1821406116}.

\bibitem[{\citenamefont{Vinkler-Aviv and Rosch}(2018)}]{Vinkler-Aviv2018.8}
\bibinfo{author}{\bibfnamefont{Y.}~\bibnamefont{Vinkler-Aviv}}
  \bibnamefont{and} \bibinfo{author}{\bibfnamefont{A.}~\bibnamefont{Rosch}},
  \bibinfo{journal}{Phys. Rev. X} \textbf{\bibinfo{volume}{8}},
  \bibinfo{pages}{031032} (\bibinfo{year}{2018}),
  \urlprefix\url{https://link.aps.org/doi/10.1103/PhysRevX.8.031032}.

\bibitem[{\citenamefont{Knolle et~al.}(2014)\citenamefont{Knolle, Kovrizhin,
  Chalker, and Moessner}}]{Knolle2014.5}
\bibinfo{author}{\bibfnamefont{J.}~\bibnamefont{Knolle}},
  \bibinfo{author}{\bibfnamefont{D.~L.} \bibnamefont{Kovrizhin}},
  \bibinfo{author}{\bibfnamefont{J.~T.} \bibnamefont{Chalker}},
  \bibnamefont{and} \bibinfo{author}{\bibfnamefont{R.}~\bibnamefont{Moessner}},
  \bibinfo{journal}{Phys. Rev. Lett.} \textbf{\bibinfo{volume}{112}},
  \bibinfo{pages}{207203} (\bibinfo{year}{2014}),
  \urlprefix\url{https://link.aps.org/doi/10.1103/PhysRevLett.112.207203}.

\bibitem[{\citenamefont{Nasu et~al.}(2016)\citenamefont{Nasu, Knolle,
  Kovrizhin, Motome, and Moessner}}]{Nasu2016.7}
\bibinfo{author}{\bibfnamefont{J.}~\bibnamefont{Nasu}},
  \bibinfo{author}{\bibfnamefont{J.}~\bibnamefont{Knolle}},
  \bibinfo{author}{\bibfnamefont{D.~L.} \bibnamefont{Kovrizhin}},
  \bibinfo{author}{\bibfnamefont{Y.}~\bibnamefont{Motome}}, \bibnamefont{and}
  \bibinfo{author}{\bibfnamefont{R.}~\bibnamefont{Moessner}},
  \bibinfo{journal}{Nature Physics} \textbf{\bibinfo{volume}{12}},
  \bibinfo{pages}{912} (\bibinfo{year}{2016}), ISSN \bibinfo{issn}{1745-2481},
  \urlprefix\url{https://doi.org/10.1038/nphys3809}.

\bibitem[{\citenamefont{Nasu et~al.}(2015)\citenamefont{Nasu, Udagawa, and
  Motome}}]{Nasu2015.9}
\bibinfo{author}{\bibfnamefont{J.}~\bibnamefont{Nasu}},
  \bibinfo{author}{\bibfnamefont{M.}~\bibnamefont{Udagawa}}, \bibnamefont{and}
  \bibinfo{author}{\bibfnamefont{Y.}~\bibnamefont{Motome}},
  \bibinfo{journal}{Phys. Rev. B} \textbf{\bibinfo{volume}{92}},
  \bibinfo{pages}{115122} (\bibinfo{year}{2015}),
  \urlprefix\url{https://link.aps.org/doi/10.1103/PhysRevB.92.115122}.

\bibitem[{\citenamefont{Nasu et~al.}(2017)\citenamefont{Nasu, Yoshitake, and
  Motome}}]{Nasu2017.9}
\bibinfo{author}{\bibfnamefont{J.}~\bibnamefont{Nasu}},
  \bibinfo{author}{\bibfnamefont{J.}~\bibnamefont{Yoshitake}},
  \bibnamefont{and} \bibinfo{author}{\bibfnamefont{Y.}~\bibnamefont{Motome}},
  \bibinfo{journal}{Phys. Rev. Lett.} \textbf{\bibinfo{volume}{119}},
  \bibinfo{pages}{127204} (\bibinfo{year}{2017}),
  \urlprefix\url{https://link.aps.org/doi/10.1103/PhysRevLett.119.127204}.

\bibitem[{\citenamefont{Motome and Nasu}(2020)}]{Motome2019.12}
\bibinfo{author}{\bibfnamefont{Y.}~\bibnamefont{Motome}} \bibnamefont{and}
  \bibinfo{author}{\bibfnamefont{J.}~\bibnamefont{Nasu}},
  \bibinfo{journal}{Journal of the Physical Society of Japan}
  \textbf{\bibinfo{volume}{89}}, \bibinfo{pages}{012002}
  (\bibinfo{year}{2020}), \eprint{https://doi.org/10.7566/JPSJ.89.012002},
  \urlprefix\url{https://doi.org/10.7566/JPSJ.89.012002}.

\bibitem[{\citenamefont{Kasahara
  et~al.}(2018{\natexlab{b}})\citenamefont{Kasahara, Sugii, Ohnishi, Shimozawa,
  Yamashita, Kurita, Tanaka, Nasu, Motome, Shibauchi et~al.}}]{Kasahara2018.5}
\bibinfo{author}{\bibfnamefont{Y.}~\bibnamefont{Kasahara}},
  \bibinfo{author}{\bibfnamefont{K.}~\bibnamefont{Sugii}},
  \bibinfo{author}{\bibfnamefont{T.}~\bibnamefont{Ohnishi}},
  \bibinfo{author}{\bibfnamefont{M.}~\bibnamefont{Shimozawa}},
  \bibinfo{author}{\bibfnamefont{M.}~\bibnamefont{Yamashita}},
  \bibinfo{author}{\bibfnamefont{N.}~\bibnamefont{Kurita}},
  \bibinfo{author}{\bibfnamefont{H.}~\bibnamefont{Tanaka}},
  \bibinfo{author}{\bibfnamefont{J.}~\bibnamefont{Nasu}},
  \bibinfo{author}{\bibfnamefont{Y.}~\bibnamefont{Motome}},
  \bibinfo{author}{\bibfnamefont{T.}~\bibnamefont{Shibauchi}},
  \bibnamefont{et~al.}, \bibinfo{journal}{Phys. Rev. Lett.}
  \textbf{\bibinfo{volume}{120}}, \bibinfo{pages}{217205}
  (\bibinfo{year}{2018}{\natexlab{b}}),
  \urlprefix\url{https://link.aps.org/doi/10.1103/PhysRevLett.120.217205}.

\bibitem[{\citenamefont{Yamada et~al.}(2017)\citenamefont{Yamada, Fujita, and
  Oshikawa}}]{Yamada2017.8}
\bibinfo{author}{\bibfnamefont{M.~G.} \bibnamefont{Yamada}},
  \bibinfo{author}{\bibfnamefont{H.}~\bibnamefont{Fujita}}, \bibnamefont{and}
  \bibinfo{author}{\bibfnamefont{M.}~\bibnamefont{Oshikawa}},
  \bibinfo{journal}{Phys. Rev. Lett.} \textbf{\bibinfo{volume}{119}},
  \bibinfo{pages}{057202} (\bibinfo{year}{2017}),
  \urlprefix\url{https://link.aps.org/doi/10.1103/PhysRevLett.119.057202}.

\bibitem[{\citenamefont{Gohlke et~al.}(2017)\citenamefont{Gohlke, Verresen,
  Moessner, and Pollmann}}]{Gohlke2017.10}
\bibinfo{author}{\bibfnamefont{M.}~\bibnamefont{Gohlke}},
  \bibinfo{author}{\bibfnamefont{R.}~\bibnamefont{Verresen}},
  \bibinfo{author}{\bibfnamefont{R.}~\bibnamefont{Moessner}}, \bibnamefont{and}
  \bibinfo{author}{\bibfnamefont{F.}~\bibnamefont{Pollmann}},
  \bibinfo{journal}{Phys. Rev. Lett.} \textbf{\bibinfo{volume}{119}},
  \bibinfo{pages}{157203} (\bibinfo{year}{2017}),
  \urlprefix\url{https://link.aps.org/doi/10.1103/PhysRevLett.119.157203}.

\bibitem[{\citenamefont{Cookmeyer and Moore}(2023)}]{Cookmeyer2023.6}
\bibinfo{author}{\bibfnamefont{T.}~\bibnamefont{Cookmeyer}} \bibnamefont{and}
  \bibinfo{author}{\bibfnamefont{J.~E.} \bibnamefont{Moore}},
  \bibinfo{journal}{Phys. Rev. B} \textbf{\bibinfo{volume}{107}},
  \bibinfo{pages}{224428} (\bibinfo{year}{2023}),
  \urlprefix\url{https://link.aps.org/doi/10.1103/PhysRevB.107.224428}.

\bibitem[{\citenamefont{Go et~al.}(2019)\citenamefont{Go, Jung, and
  Moon}}]{Go2019.4}
\bibinfo{author}{\bibfnamefont{A.}~\bibnamefont{Go}},
  \bibinfo{author}{\bibfnamefont{J.}~\bibnamefont{Jung}}, \bibnamefont{and}
  \bibinfo{author}{\bibfnamefont{E.-G.} \bibnamefont{Moon}},
  \bibinfo{journal}{Phys. Rev. Lett.} \textbf{\bibinfo{volume}{122}},
  \bibinfo{pages}{147203} (\bibinfo{year}{2019}),
  \urlprefix\url{https://link.aps.org/doi/10.1103/PhysRevLett.122.147203}.

\bibitem[{\citenamefont{Yang et~al.}(2022)\citenamefont{Yang, Kim, Choi, Lee,
  Lin, Ma, Kratochv\'{\i}lov\'a, Proschek, Moon, Lee et~al.}}]{Yang2022.8}
\bibinfo{author}{\bibfnamefont{H.}~\bibnamefont{Yang}},
  \bibinfo{author}{\bibfnamefont{C.}~\bibnamefont{Kim}},
  \bibinfo{author}{\bibfnamefont{Y.}~\bibnamefont{Choi}},
  \bibinfo{author}{\bibfnamefont{J.~H.} \bibnamefont{Lee}},
  \bibinfo{author}{\bibfnamefont{G.}~\bibnamefont{Lin}},
  \bibinfo{author}{\bibfnamefont{J.}~\bibnamefont{Ma}},
  \bibinfo{author}{\bibfnamefont{M.}~\bibnamefont{Kratochv\'{\i}lov\'a}},
  \bibinfo{author}{\bibfnamefont{P.}~\bibnamefont{Proschek}},
  \bibinfo{author}{\bibfnamefont{E.-G.} \bibnamefont{Moon}},
  \bibinfo{author}{\bibfnamefont{K.~H.} \bibnamefont{Lee}},
  \bibnamefont{et~al.}, \bibinfo{journal}{Phys. Rev. B}
  \textbf{\bibinfo{volume}{106}}, \bibinfo{pages}{L081116}
  (\bibinfo{year}{2022}),
  \urlprefix\url{https://link.aps.org/doi/10.1103/PhysRevB.106.L081116}.

\bibitem[{\citenamefont{Hal{\'a}sz}(2025)}]{Halasz.G.B.2025.05}
\bibinfo{author}{\bibfnamefont{G.~B.} \bibnamefont{Hal{\'a}sz}},
  \bibinfo{journal}{arXiv preprint arXiv:2505.03879}  (\bibinfo{year}{2025}).

\bibitem[{\citenamefont{Zhang et~al.}(2025)\citenamefont{Zhang, Halasz, Ghosh,
  Jesse, Ward, Tennant, McGuire, and Yan}}]{Zhang.H.2025.05}
\bibinfo{author}{\bibfnamefont{H.}~\bibnamefont{Zhang}},
  \bibinfo{author}{\bibfnamefont{G.~B.} \bibnamefont{Halasz}},
  \bibinfo{author}{\bibfnamefont{S.}~\bibnamefont{Ghosh}},
  \bibinfo{author}{\bibfnamefont{S.}~\bibnamefont{Jesse}},
  \bibinfo{author}{\bibfnamefont{T.~Z.} \bibnamefont{Ward}},
  \bibinfo{author}{\bibfnamefont{D.~A.} \bibnamefont{Tennant}},
  \bibinfo{author}{\bibfnamefont{M.}~\bibnamefont{McGuire}}, \bibnamefont{and}
  \bibinfo{author}{\bibfnamefont{J.}~\bibnamefont{Yan}},
  \bibinfo{journal}{arXiv preprint arXiv:2505.05417}  (\bibinfo{year}{2025}).

\bibitem[{\citenamefont{Zhang et~al.}(2023)\citenamefont{Zhang, Wilke, and
  Kim}}]{Zhang.E.Z.2023.5}
\bibinfo{author}{\bibfnamefont{E.~Z.} \bibnamefont{Zhang}},
  \bibinfo{author}{\bibfnamefont{R.~H.} \bibnamefont{Wilke}}, \bibnamefont{and}
  \bibinfo{author}{\bibfnamefont{Y.~B.} \bibnamefont{Kim}},
  \bibinfo{journal}{Phys. Rev. B} \textbf{\bibinfo{volume}{107}},
  \bibinfo{pages}{184418} (\bibinfo{year}{2023}),
  \urlprefix\url{https://link.aps.org/doi/10.1103/PhysRevB.107.184418}.

\bibitem[{\citenamefont{Halloran et~al.}(2023)\citenamefont{Halloran,
  Desrochers, Zhang, Chen, Chern, Xu, Winn, Graves-Brook, Stone, Kolesnikov
  et~al.}}]{Halloran.T.2022.11}
\bibinfo{author}{\bibfnamefont{T.}~\bibnamefont{Halloran}},
  \bibinfo{author}{\bibfnamefont{F.}~\bibnamefont{Desrochers}},
  \bibinfo{author}{\bibfnamefont{E.~Z.} \bibnamefont{Zhang}},
  \bibinfo{author}{\bibfnamefont{T.}~\bibnamefont{Chen}},
  \bibinfo{author}{\bibfnamefont{L.~E.} \bibnamefont{Chern}},
  \bibinfo{author}{\bibfnamefont{Z.}~\bibnamefont{Xu}},
  \bibinfo{author}{\bibfnamefont{B.}~\bibnamefont{Winn}},
  \bibinfo{author}{\bibfnamefont{M.}~\bibnamefont{Graves-Brook}},
  \bibinfo{author}{\bibfnamefont{M.~B.} \bibnamefont{Stone}},
  \bibinfo{author}{\bibfnamefont{A.~I.} \bibnamefont{Kolesnikov}},
  \bibnamefont{et~al.}, \bibinfo{journal}{Proceedings of the National Academy
  of Sciences} \textbf{\bibinfo{volume}{120}}, \bibinfo{pages}{e2215509119}
  (\bibinfo{year}{2023}),
  \eprint{https://www.pnas.org/doi/pdf/10.1073/pnas.2215509119},
  \urlprefix\url{https://www.pnas.org/doi/abs/10.1073/pnas.2215509119}.

\bibitem[{\citenamefont{Fletcher et~al.}(1967)\citenamefont{Fletcher, Gardner,
  Fox, and Topping}}]{Fletcher1967.1}
\bibinfo{author}{\bibfnamefont{J.~M.} \bibnamefont{Fletcher}},
  \bibinfo{author}{\bibfnamefont{W.~E.} \bibnamefont{Gardner}},
  \bibinfo{author}{\bibfnamefont{A.~C.} \bibnamefont{Fox}}, \bibnamefont{and}
  \bibinfo{author}{\bibfnamefont{G.}~\bibnamefont{Topping}},
  \bibinfo{journal}{J. Chem. Soc. A} pp. \bibinfo{pages}{1038--1045}
  (\bibinfo{year}{1967}),
  \urlprefix\url{http://dx.doi.org/10.1039/J19670001038}.

\bibitem[{\citenamefont{Kubota et~al.}(2015)\citenamefont{Kubota, Tanaka, Ono,
  Narumi, and Kindo}}]{Kubota2015.3}
\bibinfo{author}{\bibfnamefont{Y.}~\bibnamefont{Kubota}},
  \bibinfo{author}{\bibfnamefont{H.}~\bibnamefont{Tanaka}},
  \bibinfo{author}{\bibfnamefont{T.}~\bibnamefont{Ono}},
  \bibinfo{author}{\bibfnamefont{Y.}~\bibnamefont{Narumi}}, \bibnamefont{and}
  \bibinfo{author}{\bibfnamefont{K.}~\bibnamefont{Kindo}},
  \bibinfo{journal}{Phys. Rev. B} \textbf{\bibinfo{volume}{91}},
  \bibinfo{pages}{094422} (\bibinfo{year}{2015}),
  \urlprefix\url{https://link.aps.org/doi/10.1103/PhysRevB.91.094422}.

\bibitem[{\citenamefont{Kaib et~al.}(2021)\citenamefont{Kaib, Biswas, Riedl,
  Winter, and Valent\'{\i}}}]{Kaib2021.4}
\bibinfo{author}{\bibfnamefont{D.~A.~S.} \bibnamefont{Kaib}},
  \bibinfo{author}{\bibfnamefont{S.}~\bibnamefont{Biswas}},
  \bibinfo{author}{\bibfnamefont{K.}~\bibnamefont{Riedl}},
  \bibinfo{author}{\bibfnamefont{S.~M.} \bibnamefont{Winter}},
  \bibnamefont{and}
  \bibinfo{author}{\bibfnamefont{R.}~\bibnamefont{Valent\'{\i}}},
  \bibinfo{journal}{Phys. Rev. B} \textbf{\bibinfo{volume}{103}},
  \bibinfo{pages}{L140402} (\bibinfo{year}{2021}),
  \urlprefix\url{https://link.aps.org/doi/10.1103/PhysRevB.103.L140402}.

\bibitem[{\citenamefont{Kim et~al.}(2015)\citenamefont{Kim, V., Catuneanu, and
  Kee}}]{kim_kitaev_2015}
\bibinfo{author}{\bibfnamefont{H.-S.} \bibnamefont{Kim}},
  \bibinfo{author}{\bibfnamefont{V.~S.} \bibnamefont{V.}},
  \bibinfo{author}{\bibfnamefont{A.}~\bibnamefont{Catuneanu}},
  \bibnamefont{and} \bibinfo{author}{\bibfnamefont{H.-Y.} \bibnamefont{Kee}},
  \bibinfo{journal}{Phys. Rev. B} \textbf{\bibinfo{volume}{91}},
  \bibinfo{pages}{241110} (\bibinfo{year}{2015}),
  \urlprefix\url{https://link.aps.org/doi/10.1103/PhysRevB.91.241110}.

\bibitem[{\citenamefont{Sarikurt et~al.}(2018)\citenamefont{Sarikurt, Kadioglu,
  Ersan, Vatansever, Akturk, Yuksel, Akıncı, and
  Akturk}}]{sarikurt_electronic_2018}
\bibinfo{author}{\bibfnamefont{S.}~\bibnamefont{Sarikurt}},
  \bibinfo{author}{\bibfnamefont{Y.}~\bibnamefont{Kadioglu}},
  \bibinfo{author}{\bibfnamefont{F.}~\bibnamefont{Ersan}},
  \bibinfo{author}{\bibfnamefont{E.}~\bibnamefont{Vatansever}},
  \bibinfo{author}{\bibfnamefont{O.~U.} \bibnamefont{Akturk}},
  \bibinfo{author}{\bibfnamefont{Y.}~\bibnamefont{Yuksel}},
  \bibinfo{author}{\bibfnamefont{U.}~\bibnamefont{Akıncı}}, \bibnamefont{and}
  \bibinfo{author}{\bibfnamefont{E.}~\bibnamefont{Akturk}},
  \bibinfo{journal}{Phys. Chem. Chem. Phys.} \textbf{\bibinfo{volume}{20}},
  \bibinfo{pages}{997} (\bibinfo{year}{2018}), ISSN \bibinfo{issn}{1463-9084},
  \urlprefix\url{https://pubs.rsc.org/en/content/articlelanding/2018/cp/c7cp07953b}.

\bibitem[{\citenamefont{Iyikanat et~al.}(2018)\citenamefont{Iyikanat,
  Yagmurcukardes, Senger, and Sahin}}]{iyikanat_tuning_2018}
\bibinfo{author}{\bibfnamefont{F.}~\bibnamefont{Iyikanat}},
  \bibinfo{author}{\bibfnamefont{M.}~\bibnamefont{Yagmurcukardes}},
  \bibinfo{author}{\bibfnamefont{R.~T.} \bibnamefont{Senger}},
  \bibnamefont{and} \bibinfo{author}{\bibfnamefont{H.}~\bibnamefont{Sahin}},
  \bibinfo{journal}{J. Mater. Chem. C} \textbf{\bibinfo{volume}{6}},
  \bibinfo{pages}{2019} (\bibinfo{year}{2018}), ISSN \bibinfo{issn}{2050-7534},
  \urlprefix\url{https://pubs.rsc.org/en/content/articlelanding/2018/tc/c7tc05266a}.

\bibitem[{\citenamefont{Vatansever et~al.}(2019)\citenamefont{Vatansever,
  Sarikurt, Ersan, Kadioglu, Uzengi~Akturk, Yuksel, Ataca, Akturk, and
  Akıncı}}]{vatansever_strain_2019}
\bibinfo{author}{\bibfnamefont{E.}~\bibnamefont{Vatansever}},
  \bibinfo{author}{\bibfnamefont{S.}~\bibnamefont{Sarikurt}},
  \bibinfo{author}{\bibfnamefont{F.}~\bibnamefont{Ersan}},
  \bibinfo{author}{\bibfnamefont{Y.}~\bibnamefont{Kadioglu}},
  \bibinfo{author}{\bibfnamefont{O.}~\bibnamefont{Uzengi~Akturk}},
  \bibinfo{author}{\bibfnamefont{Y.}~\bibnamefont{Yuksel}},
  \bibinfo{author}{\bibfnamefont{C.}~\bibnamefont{Ataca}},
  \bibinfo{author}{\bibfnamefont{E.}~\bibnamefont{Akturk}}, \bibnamefont{and}
  \bibinfo{author}{\bibfnamefont{U.}~\bibnamefont{Akıncı}},
  \bibinfo{journal}{J. Appl. Phys.} \textbf{\bibinfo{volume}{125}},
  \bibinfo{pages}{083903} (\bibinfo{year}{2019}), ISSN
  \bibinfo{issn}{0021-8979}, \urlprefix\url{https://doi.org/10.1063/1.5078713}.

\bibitem[{\citenamefont{Liu et~al.}(2023)\citenamefont{Liu, Yang, Wang, Lu, Ma,
  and Wu}}]{liu_contrasting_2023}
\bibinfo{author}{\bibfnamefont{L.}~\bibnamefont{Liu}},
  \bibinfo{author}{\bibfnamefont{K.}~\bibnamefont{Yang}},
  \bibinfo{author}{\bibfnamefont{G.}~\bibnamefont{Wang}},
  \bibinfo{author}{\bibfnamefont{D.}~\bibnamefont{Lu}},
  \bibinfo{author}{\bibfnamefont{Y.}~\bibnamefont{Ma}}, \bibnamefont{and}
  \bibinfo{author}{\bibfnamefont{H.}~\bibnamefont{Wu}}, \bibinfo{journal}{Phys.
  Rev. B} \textbf{\bibinfo{volume}{107}}, \bibinfo{pages}{165134}
  (\bibinfo{year}{2023}),
  \urlprefix\url{https://link.aps.org/doi/10.1103/PhysRevB.107.165134}.

\bibitem[{\citenamefont{Samanta et~al.}(2024)\citenamefont{Samanta, Hong, and
  Kim}}]{samanta_electronic_2024}
\bibinfo{author}{\bibfnamefont{S.}~\bibnamefont{Samanta}},
  \bibinfo{author}{\bibfnamefont{D.}~\bibnamefont{Hong}}, \bibnamefont{and}
  \bibinfo{author}{\bibfnamefont{H.-S.} \bibnamefont{Kim}},
  \bibinfo{journal}{Nanomaterials} \textbf{\bibinfo{volume}{14}},
  \bibinfo{pages}{9} (\bibinfo{year}{2024}), ISSN \bibinfo{issn}{2079-4991},
  \urlprefix\url{https://www.mdpi.com/2079-4991/14/1/9}.

\bibitem[{\citenamefont{Sugano et~al.}(1970)\citenamefont{Sugano, Tanabe, and
  Kamimura}}]{Kanamori}
\bibinfo{author}{\bibfnamefont{S.}~\bibnamefont{Sugano}},
  \bibinfo{author}{\bibfnamefont{Y.}~\bibnamefont{Tanabe}}, \bibnamefont{and}
  \bibinfo{author}{\bibfnamefont{H.}~\bibnamefont{Kamimura}},
  \emph{\bibinfo{title}{Multiplets of Transition-Metal Ions in Crystals}},
  vol.~\bibinfo{volume}{33} of \emph{\bibinfo{series}{Pure and Applied
  Physics}} (\bibinfo{publisher}{Academic Press}, \bibinfo{address}{New York},
  \bibinfo{year}{1970}).

\bibitem[{\citenamefont{Noh et~al.}(2024)\citenamefont{Noh, Hwang, and
  Moon}}]{Noh2024.5}
\bibinfo{author}{\bibfnamefont{P.}~\bibnamefont{Noh}},
  \bibinfo{author}{\bibfnamefont{K.}~\bibnamefont{Hwang}}, \bibnamefont{and}
  \bibinfo{author}{\bibfnamefont{E.-G.} \bibnamefont{Moon}},
  \bibinfo{journal}{Phys. Rev. B} \textbf{\bibinfo{volume}{109}},
  \bibinfo{pages}{L201105} (\bibinfo{year}{2024}),
  \urlprefix\url{https://link.aps.org/doi/10.1103/PhysRevB.109.L201105}.

\bibitem[{\citenamefont{Hwang et~al.}(2022)\citenamefont{Hwang, Go, Seong,
  Shibauchi, and Moon}}]{Hwang2022.1}
\bibinfo{author}{\bibfnamefont{K.}~\bibnamefont{Hwang}},
  \bibinfo{author}{\bibfnamefont{A.}~\bibnamefont{Go}},
  \bibinfo{author}{\bibfnamefont{J.~H.} \bibnamefont{Seong}},
  \bibinfo{author}{\bibfnamefont{T.}~\bibnamefont{Shibauchi}},
  \bibnamefont{and} \bibinfo{author}{\bibfnamefont{E.-G.} \bibnamefont{Moon}},
  \bibinfo{journal}{Nature Communications} \textbf{\bibinfo{volume}{13}},
  \bibinfo{pages}{323} (\bibinfo{year}{2022}), ISSN \bibinfo{issn}{2041-1723},
  \urlprefix\url{https://doi.org/10.1038/s41467-021-27943-9}.

\bibitem[{\citenamefont{Imamura et~al.}(2024)\citenamefont{Imamura, Suetsugu,
  Mizukami, Yoshida, Hashimoto, Ohtsuka, Kasahara, Kurita, Tanaka, Noh
  et~al.}}]{Imamura2023.5}
\bibinfo{author}{\bibfnamefont{K.}~\bibnamefont{Imamura}},
  \bibinfo{author}{\bibfnamefont{S.}~\bibnamefont{Suetsugu}},
  \bibinfo{author}{\bibfnamefont{Y.}~\bibnamefont{Mizukami}},
  \bibinfo{author}{\bibfnamefont{Y.}~\bibnamefont{Yoshida}},
  \bibinfo{author}{\bibfnamefont{K.}~\bibnamefont{Hashimoto}},
  \bibinfo{author}{\bibfnamefont{K.}~\bibnamefont{Ohtsuka}},
  \bibinfo{author}{\bibfnamefont{Y.}~\bibnamefont{Kasahara}},
  \bibinfo{author}{\bibfnamefont{N.}~\bibnamefont{Kurita}},
  \bibinfo{author}{\bibfnamefont{H.}~\bibnamefont{Tanaka}},
  \bibinfo{author}{\bibfnamefont{P.}~\bibnamefont{Noh}}, \bibnamefont{et~al.},
  \bibinfo{journal}{Science Advances} \textbf{\bibinfo{volume}{10}},
  \bibinfo{pages}{eadk3539} (\bibinfo{year}{2024}),
  \eprint{https://www.science.org/doi/pdf/10.1126/sciadv.adk3539},
  \urlprefix\url{https://www.science.org/doi/abs/10.1126/sciadv.adk3539}.

\bibitem[{\citenamefont{Kim et~al.}(2024)\citenamefont{Kim, Park, Samanta,
  Choi, Kang, Seo, Ji, Noh, Cho, Yoo et~al.}}]{Kim2024.7}
\bibinfo{author}{\bibfnamefont{G.-H.} \bibnamefont{Kim}},
  \bibinfo{author}{\bibfnamefont{M.}~\bibnamefont{Park}},
  \bibinfo{author}{\bibfnamefont{S.}~\bibnamefont{Samanta}},
  \bibinfo{author}{\bibfnamefont{U.}~\bibnamefont{Choi}},
  \bibinfo{author}{\bibfnamefont{B.}~\bibnamefont{Kang}},
  \bibinfo{author}{\bibfnamefont{U.}~\bibnamefont{Seo}},
  \bibinfo{author}{\bibfnamefont{G.}~\bibnamefont{Ji}},
  \bibinfo{author}{\bibfnamefont{S.}~\bibnamefont{Noh}},
  \bibinfo{author}{\bibfnamefont{D.-Y.} \bibnamefont{Cho}},
  \bibinfo{author}{\bibfnamefont{J.-W.} \bibnamefont{Yoo}},
  \bibnamefont{et~al.}, \bibinfo{journal}{Science Advances}
  \textbf{\bibinfo{volume}{10}}, \bibinfo{pages}{eadn8694}
  (\bibinfo{year}{2024}),
  \eprint{https://www.science.org/doi/pdf/10.1126/sciadv.adn8694},
  \urlprefix\url{https://www.science.org/doi/abs/10.1126/sciadv.adn8694}.

\bibitem[{\citenamefont{Perdew et~al.}(1996)\citenamefont{Perdew, Burke, and
  Ernzerhof}}]{perdew_generalized_1996}
\bibinfo{author}{\bibfnamefont{J.~P.} \bibnamefont{Perdew}},
  \bibinfo{author}{\bibfnamefont{K.}~\bibnamefont{Burke}}, \bibnamefont{and}
  \bibinfo{author}{\bibfnamefont{M.}~\bibnamefont{Ernzerhof}},
  \bibinfo{journal}{Phys. Rev. Lett.} \textbf{\bibinfo{volume}{77}},
  \bibinfo{pages}{3865} (\bibinfo{year}{1996}),
  \urlprefix\url{https://link.aps.org/doi/10.1103/PhysRevLett.77.3865}.

\bibitem[{\citenamefont{Kresse and
  Furthmüller}(1996{\natexlab{a}})}]{kresse_efficiency_1996}
\bibinfo{author}{\bibfnamefont{G.}~\bibnamefont{Kresse}} \bibnamefont{and}
  \bibinfo{author}{\bibfnamefont{J.}~\bibnamefont{Furthmüller}},
  \bibinfo{journal}{Computational Materials Science}
  \textbf{\bibinfo{volume}{6}}, \bibinfo{pages}{15}
  (\bibinfo{year}{1996}{\natexlab{a}}), ISSN \bibinfo{issn}{0927-0256},
  \urlprefix\url{https://www.sciencedirect.com/science/article/pii/0927025696000080}.

\bibitem[{\citenamefont{Kresse and
  Furthmüller}(1996{\natexlab{b}})}]{kresse_efficient_1996}
\bibinfo{author}{\bibfnamefont{G.}~\bibnamefont{Kresse}} \bibnamefont{and}
  \bibinfo{author}{\bibfnamefont{J.}~\bibnamefont{Furthmüller}},
  \bibinfo{journal}{Phys. Rev. B} \textbf{\bibinfo{volume}{54}},
  \bibinfo{pages}{11169} (\bibinfo{year}{1996}{\natexlab{b}}),
  \urlprefix\url{https://link.aps.org/doi/10.1103/PhysRevB.54.11169}.

\bibitem[{\citenamefont{Ozaki}(2003)}]{ozaki_variationally_2003}
\bibinfo{author}{\bibfnamefont{T.}~\bibnamefont{Ozaki}},
  \bibinfo{journal}{Phys. Rev. B} \textbf{\bibinfo{volume}{67}},
  \bibinfo{pages}{155108} (\bibinfo{year}{2003}),
  \urlprefix\url{https://link.aps.org/doi/10.1103/PhysRevB.67.155108}.

\bibitem[{\citenamefont{Kim et~al.}(2014)\citenamefont{Kim, Kim, and
  Han}}]{kim_electronic_2014}
\bibinfo{author}{\bibfnamefont{K.-H.} \bibnamefont{Kim}},
  \bibinfo{author}{\bibfnamefont{H.-S.} \bibnamefont{Kim}}, \bibnamefont{and}
  \bibinfo{author}{\bibfnamefont{M.~J.} \bibnamefont{Han}},
  \bibinfo{journal}{J. Phys.: Condens. Matter} \textbf{\bibinfo{volume}{26}},
  \bibinfo{pages}{185501} (\bibinfo{year}{2014}), ISSN
  \bibinfo{issn}{0953-8984},
  \urlprefix\url{https://dx.doi.org/10.1088/0953-8984/26/18/185501}.

\bibitem[{\citenamefont{Dudarev et~al.}(1998)\citenamefont{Dudarev, Botton,
  Savrasov, Humphreys, and Sutton}}]{dudarev_electron-energy-loss_1998}
\bibinfo{author}{\bibfnamefont{S.~L.} \bibnamefont{Dudarev}},
  \bibinfo{author}{\bibfnamefont{G.~A.} \bibnamefont{Botton}},
  \bibinfo{author}{\bibfnamefont{S.~Y.} \bibnamefont{Savrasov}},
  \bibinfo{author}{\bibfnamefont{C.~J.} \bibnamefont{Humphreys}},
  \bibnamefont{and} \bibinfo{author}{\bibfnamefont{A.~P.}
  \bibnamefont{Sutton}}, \bibinfo{journal}{Phys. Rev. B}
  \textbf{\bibinfo{volume}{57}}, \bibinfo{pages}{1505} (\bibinfo{year}{1998}),
  \urlprefix\url{https://link.aps.org/doi/10.1103/PhysRevB.57.1505}.

\bibitem[{\citenamefont{Han et~al.}(2006)\citenamefont{Han, Ozaki, and
  Yu}}]{han_mathrmon_2006}
\bibinfo{author}{\bibfnamefont{M.~J.} \bibnamefont{Han}},
  \bibinfo{author}{\bibfnamefont{T.}~\bibnamefont{Ozaki}}, \bibnamefont{and}
  \bibinfo{author}{\bibfnamefont{J.}~\bibnamefont{Yu}}, \bibinfo{journal}{Phys.
  Rev. B} \textbf{\bibinfo{volume}{73}}, \bibinfo{pages}{045110}
  (\bibinfo{year}{2006}),
  \urlprefix\url{https://link.aps.org/doi/10.1103/PhysRevB.73.045110}.

\bibitem[{\citenamefont{Pizzi et~al.}(2020)\citenamefont{Pizzi, Vitale, Arita,
  Blügel, Freimuth, Géranton, Gibertini, Gresch, Johnson, Koretsune
  et~al.}}]{pizzi_wannier90_2020}
\bibinfo{author}{\bibfnamefont{G.}~\bibnamefont{Pizzi}},
  \bibinfo{author}{\bibfnamefont{V.}~\bibnamefont{Vitale}},
  \bibinfo{author}{\bibfnamefont{R.}~\bibnamefont{Arita}},
  \bibinfo{author}{\bibfnamefont{S.}~\bibnamefont{Blügel}},
  \bibinfo{author}{\bibfnamefont{F.}~\bibnamefont{Freimuth}},
  \bibinfo{author}{\bibfnamefont{G.}~\bibnamefont{Géranton}},
  \bibinfo{author}{\bibfnamefont{M.}~\bibnamefont{Gibertini}},
  \bibinfo{author}{\bibfnamefont{D.}~\bibnamefont{Gresch}},
  \bibinfo{author}{\bibfnamefont{C.}~\bibnamefont{Johnson}},
  \bibinfo{author}{\bibfnamefont{T.}~\bibnamefont{Koretsune}},
  \bibnamefont{et~al.}, \bibinfo{journal}{J. Phys.: Condens. Matter}
  \textbf{\bibinfo{volume}{32}}, \bibinfo{pages}{165902}
  (\bibinfo{year}{2020}), ISSN \bibinfo{issn}{0953-8984},
  \urlprefix\url{https://dx.doi.org/10.1088/1361-648X/ab51ff}.

\end{thebibliography}

\end{document}